\documentclass[12pt]{article}
\usepackage[linktocpage]{hyperref}
\usepackage[noadjust]{cite}
\usepackage{float}
\usepackage{amsfonts}
\usepackage{comment}
\usepackage{csquotes}
\usepackage{subcaption}
\usepackage{amsmath}
\usepackage{amssymb}
\usepackage{latexsym}
\usepackage{graphicx}
\usepackage[english]{babel}
\usepackage[font={small}]{caption}
\usepackage{extarrows}
\usepackage{fancybox}
\usepackage{mathtools}

\usepackage{url}
\usepackage{enumerate}
\usepackage[shortlabels]{enumitem}
\usepackage{ulem}
\usepackage{tikz}

\begin{document}
\input epsf

\def\p{\partial}
\def\h{{\frac12}}
\def\be{\begin{equation}}
\def\bea{\begin{eqnarray}}
\def\ee{\end{equation}}
\def\eea{\end{eqnarray}}
\def\d{\partial}
\def\la{\lambda}
\def\eps{\epsilon}
\def\b{\bigskip}
\def\m{\medskip}

\newcommand{\newsection}[1]{\section{#1} \setcounter{equation}{0}}

\def\q{\quad}
\def\t{\tilde}
\def\r{\rightarrow}
\def\nn{\nonumber\\}

\newcommand{\Romk}[1]{|0^-\rangle_R^{(#1)}}
\newcommand{\Romb}[1]{{}^{\,(#1)}_{\ \,R}\langle0^-|}
\newcommand{\Ropk}[1]{|0^+\rangle_R^{(#1)}}
\newcommand{\Ropb}[1]{{}^{\,(#1)}_{\ \,R}\langle0^+|}
\newcommand{\Rok}[1]{|0\rangle_R^{(#1)}}
\newcommand{\Rob}[1]{{}^{\,(#1)}_{\ \,R}\langle0|}
\newcommand{\Rotk}[1]{|\tilde{0}\rangle_R^{(#1)}}
\newcommand{\Rotb}[1]{{}^{\,(#1)}_{\ \,R}\langle\tilde{0}|}

\newcommand{\Robmk}[1]{|\bar0^-\rangle_R^{(#1)}}
\newcommand{\Robmb}[1]{{}^{\,(#1)}_{\ \,R}\langle\bar0^-|}
\newcommand{\Robpk}[1]{|\bar0^+\rangle_R^{(#1)}}
\newcommand{\Robpb}[1]{{}^{\,(#1)}_{\ \,R}\langle\bar0^+|}
\newcommand{\Robk}[1]{|\bar0\rangle_R^{(#1)}}
\newcommand{\Robb}[1]{{}^{\,(#1)}_{\ \,R}\langle\bar0|}
\newcommand{\Rotbk}[1]{|\bar\tilde{0}\rangle_R^{(#1)}}
\newcommand{\Rotbb}[1]{{}^{\,(#1)}_{\ \,R}\langle\bar\tilde{0}|}

\newcommand{\Roobmk}[1]{|0^-,\bar0^-\rangle_R^{(#1)}}
\newcommand{\Roobmb}[1]{{}^{\,(#1)}_{\ \,R}\langle0^,\bar0^-|}

\def\NSo{0_{\scriptscriptstyle{N\!S}}}
\newcommand{\NSok}[1]{|0\rangle_{\scriptscriptstyle{N\!S}}^{(#1)}}
\newcommand{\NSob}[1]{{}^{\ \,(#1)}_{\ \;\scriptscriptstyle{N\!S}}\langle0|}

\def\phiR{\phi_R}
\def\phiNS{\phi_{\scriptscriptstyle{N\!S}}}
\newcommand{\phiRk}[1]{|\phi\rangle_R^{(#1)}}
\newcommand{\phiRb}[1]{{}^{\,(#1)}_{\ \,R}\langle\phi|}
\newcommand{\phiNSk}[1]{|\phi\rangle_{\scriptscriptstyle{N\!S}}^{(#1)}}
\newcommand{\phiNSb}[1]{{}^{\ \,(#1)}_{\ \;\scriptscriptstyle{N\!S}}\langle\phi|}

\newcommand{\MyRed}{\color [rgb]{0.8,0,0}}
\newcommand{\MyGreen}{\color [rgb]{0,0.7,0}}
\newcommand{\MyBlue}{\color [rgb]{0,0,0.8}}
\newcommand{\MyBrown}{\color [rgb]{0.8,0.4,0.1}}
\newcommand{\MyPurple}{\color [rgb]{0.6,0.0,0.6}}
\def\MH#1{{\MyRed [MH: #1]}}
\def\SM#1{{\MyPurple [SM: #1]}}   
\def\BG#1{{\MyBlue [BG: #1]}}
\def\MM#1{{\MyGreen [MM: #1]}}

\newcommand\blfootnote[1]{%
  \begingroup
  \renewcommand\thefootnote{}\footnote{#1}%
  \addtocounter{footnote}{-1}%
  \endgroup
}

\begin{flushright}
\end{flushright}
\hfill
\vspace{18pt}
\begin{center}
{\Large 
\textbf{Universal lifting in the D1-D5 CFT
}}

\end{center}

\vspace{10mm}
\begin{center}
{\textsl{Bin Guo$^{a}$}\blfootnote{${}^{a}$bin.guo@ipht.fr}\textsl{, Marcel R. R. Hughes$^{b}$}\blfootnote{${}^{b}$hughes.2059@osu.edu}\textsl{, Samir D. Mathur${}^{c}$}\blfootnote{${}^c$mathur.16@osu.edu}\textsl{ and Madhur Mehta${}^{d}$}\blfootnote{${}^d$mehta.493@osu.edu}
\\}

\vspace{10mm}

\textit{\small ${}^a$  Institut de Physique Th\'eorique,
Universit\'e Paris-Saclay,\\
CNRS, CEA, 91191, Gif-sur-Yvette, France} \\  \vspace{6pt}

\textit{\small ${}^{b,c,d}$ Department of Physics, The Ohio State University,\\ Columbus,
OH 43210, USA} \\  \vspace{6pt}

\end{center}

\vspace{12pt}
\begin{center}
\textbf{Abstract}
\end{center}

\vspace{4pt} {\small
\noindent
We consider D1-D5-P states in the untwisted sector of the D1-D5 orbifold CFT where one copy of the seed CFT has been excited with a left-moving superconformal primary. Despite being BPS at the orbifold point, such states can `lift' as the theory is deformed away from this point in moduli space. We compute this lifting at second order in the deformation parameter for arbitrary left-moving dimension $h$ of this class of states. This result displays an interesting universality since the lifting does not depend on the details of the superconformal primary; it depends only on the dimension. In the large-dimension limit the lift scales as $\sqrt{h}\,$; it is observed that such scaling appears to be a universal property of the lift of D1-D5-P states.

\vspace{1cm}

\thispagestyle{empty}

\setcounter{footnote}{0}
\setcounter{page}{0}
\newpage

\tableofcontents

\setcounter{page}{1}

\numberwithin{equation}{section}

\section{Introduction}\label{secintro}
\normalem

The bound states of D1-brane, D5-brane and momentum (P) charges in string theory have proven to be very useful examples of black hole microstates. In classical gravity, there exists a solution with these same charges and with a Bekenstein entropy of $S_{\rm{bek}}=A/4G$. The number of BPS bound states with these charges should give the microscopic entropy of the system. An index counting these states was computed in \cite{Strominger:1996sh} for type IIB string theory compactified on $K3\times S^1$ and in \cite{Maldacena:1999bp} for IIB compactified on $T^4\times S^1$. In each case the leading order behavior of the index matches the Bekenstein entropy.

More detailed considerations of the microstates reveal some very interesting questions. The D1-D5 conformal field theory (CFT) is conjectured to have a so-called `orbifold point' in its  moduli space, at which the CFT is described by a ($1+1$)-dimensional sigma model with an orbifold target space \cite{Vafa:1995bm,Dijkgraaf:1998gf,Seiberg:1999xz,Larsen:1999uk,Arutyunov:1997gi,Arutyunov:1997gt,Jevicki:1998bm,David:2002wn}; this is the analogue of free super Yang-Mills (SYM) in the D3-brane case. At this orbifold point, any state of the CFT with only left-moving (or equally, only right-moving) excitations is a BPS state -- some fraction of the $\mathcal{N}=4$ supersymmetry of the theory is preserved. As the theory is deformed away from the orbifold point, towards the part of the moduli space for which there is a dual semiclassical gravity description, some of the short multiplets of these BPS states join into long multiplets and `lift'. That is, they gain an anomalous contribution to their conformal dimension. Therefore, only the states that remain unlifted away from the orbifold point can be globally BPS and thus contribute to the index. This situation raises a natural question: which states lift and by how much? The question is of interest for two main reasons.

Firstly, the fuzzball program\footnote{For a window into the current state of the fuzzball and microstate geometry programs see the general and technical reviews \cite{Bena:2022ldq} and \cite{Bena:2022rna} respectively.} has been able to construct large classes of microstates of black holes and in the process provide a heuristic map between these gravity states and states of the D1-D5 CFT at the orbifold point \cite{Lunin:2001jy,Mathur:2005zp,Kanitscheider:2007wq,Bena:2007kg,Chowdhury:2010ct,Shigemori:2020yuo}. Since different gravitational microstates have different dynamical properties, it is interesting to know what class of states in the orbifold theory end up being unlifted and therefore can be microstates of the extremal hole.

Secondly, one encounters a puzzling feature when considering the actual computation of the lift. Consider the Neveu-Schwarz (NS) sector of the CFT where the ground state $|0\rangle_{\scriptscriptstyle{N\!S}}$ describes global AdS${}_3\times S^3\times T^4$ in the gravity theory. In \cite{Gava:2002xb}, Gava and Narain computed the lift of low-energy states in the CFT and found that almost all were lifted; this agrees with the gravity theory where the only unlifted low-lying states describe supergravity quanta, while the vast majority of excitations describe string states that are lifted up to the string scale. However, we know that at high enough excitation energies we will reach a black hole phase where a large number of states must be {\it unlifted} in order to account for the large result of index calculations. Thus the puzzling question is: what changes in the lifting computation when we look at states with energies above the black hole threshold? An answer to this question would also address the first question above, \textit{i.e.} it would tell us about the nature of microstates that contribute to the extremal hole's entropy.

Thus one would like to understand the pattern of lifting in the D1-D5 CFT, something that we do not have a clear understanding of at the present time. Lifting calculations for several different families of states have been performed already, which we briefly summarise. In \cite{Gaberdiel:2015uca}, the lift was computed for higher-spin states of the enhanced symmetry algebra found at the orbifold point. The pattern of their anomalous dimensions was studied and Regge trajectories were identified for these lifted states. In \cite{Hampton:2018ygz}, states with some copies excited above the NS vacuum with the specific excitation $J^+_{-(2m-1)}\cdots J^+_{-3}J^+_{-1}$ were considered and their lift computed. The lift for all low-level states with right-moving Ramond ground state was then computed for the case of $N=2$ in \cite{Guo:2019ady,Guo:2020gxm}. It was found that all such states that could lift\footnote{States that are BPS at the orbifold point are in short multiplets of the symmetry algebra. In order to lift, four short multiplets must join into a long multiplet of the deformed theory. The existence of suitable short multiplets is what is meant by a state being allowed to lift.}, were indeed lifted. The series of papers \cite{Lima:2020boh,Lima:2020kek,Lima:2020nnx,Lima:2020urq,Lima:2021wrz,Lima:2021xqj,Lima:2022cnq} have variously considered lifting problems associated with different Ramond sector states and composites thereof. The paper \cite{Benjamin:2021zkn} systematically computed the lifting for general $N$ of untwisted-sector $1/4$-BPS states in $(h,j)=(1,0)$ left-moving long multiplets and $\bar{j}$ right-moving short multiplets.

In the present paper, we will compute the lift for another family of states. Consider states of the D1-D5 CFT in the untwisted sector. Let all $N$ copies of the $c=6$ seed CFT be in the ground state, except for one copy which is excited to a superconformal primary with NS-sector dimension $h$ and charge $m$. There is no lift to first order in the deformation parameter $\lambda$. For this particular class of states the only lifting to consider turns out to be for states with $m=0$, for which we compute the lift (\textit{i.e.} the anomalous dimension) $E^{(2)}$ at second order in $\lambda$. Interestingly, the answer does not depend on the particular choice of state, only on the dimension $h$
\begin{equation}
    E^{(2)}_{h} = (N-1)\frac{\lambda^2\pi^2}{2^{2h-1}} \frac{\Gamma(2h)}{\Gamma(h)^2} \ .
\end{equation}
We emphasise that knowledge of the form of the state is not required to find this lift; in all previous D1-D5 lifting examples, the precise details of the considered state are used.

The paper is structured as follows. Section~\ref{secD1D5} briefly describes some key features of the D1-D5 CFT; its symmetries, spectral flow, Ramond-sector ground states, deformations away from the orbifold point, and the associated lifting of states. In Section~\ref{sec.main} we detail the computation of the second order lift in energy for untwisted-sector states with one copy excited above the ground state by a superconformal primary. This is followed by a check of the resulting lift and a brief analysis of the large-dimension limit. We conclude in Section~\ref{sec.conc} with a discussion of future directions and questions. Appendix~\ref{app_cft} provides details of the $\mathcal{N}=4$ superconformal algebra and our conventions for Hermitian conjugation, while Appendix~\ref{app.B} derives certain symmetry properties of contributions to the lift from different configurations of copies. In Appendix~\ref{app.SF} the transformation of an inserted operator under spectral flow is derived, both for the cylinder and the plane.

\section{The D1-D5 CFT} \label{secD1D5}

We consider type IIB string theory compactified as
\be
    M_{9,1}\r M_{4,1}\times S^1\times T^4 \ ,
\ee
with $n_1$ D1-branes wrapped on the $S^1$ and $n_5$ D5-branes wrapped on the $S^1\times T^4$. The bound states of these branes in the IR generate the D1-D5 CFT, which is a (1+1)-dimensional field theory living on the cylinder made from the $S^1$ and time directions. This theory is believed to have an orbifold point in its moduli space, where we have a description in terms of
\be
    N = n_1 n_5 \ ,
\ee
copies of a seed $c=6$ free CFT. This free CFT is made up of $4$ free bosons and $4$ free fermions in the left-moving sector and likewise in the right-moving sector. The free fields are subject to an orbifold symmetry generated by the group of permutations $S_N$, leading to the Hilbert space factoring into various twisted sectors labelled by $1\leq k\leq N$. Each of these sectors effectively describes a CFT on a $k$-wound circle $S^1$ (sometimes referred to as a component string). This orbifold point has been shown to be dual to a tensionless string in an AdS${}_3\times S^3\times T^4$ background with one unit of NS-NS flux~\cite{Gaberdiel:2018rqv,Eberhardt:2018ouy,Eberhardt:2019ywk,Eberhardt:2020akk}, with extensions of these ideas found, for example, in \cite{Eberhardt:2019qcl,Dei:2019osr}. It is thought that this free point can be deformed in the direction of a strong coupling regime, at which the theory would have a dual semiclassical (super)gravity description.

\subsection{Symmetries of the CFT}

At a general point in the moduli space the D1-D5 CFT has $(4,4)$ supersymmetry~\cite{Schwimmer:1986mf,Sevrin:1988ew}, which means that we have $\mathcal{N}=4$ supersymmetry in both the left- and right-moving sectors. This leads to a superconformal ${\cal N}=4$ symmetry in both the left and right sectors, with chiral algebra generators $L_{n}, G^{\alpha}_{\dot A,r}, J^a_n$ for the left movers associated with the stress-energy tensor, the supercurrents and the $\mathfrak{su}(2)$ R-currents. The right-moving sector has the analogous generators $\bar L_{n}, \bar G^{\bar \alpha}_{\dot A,r}, \bar J^a_n$. This small $\mathcal N=4$ superconformal algebra and our conventions are outlined in Appendix~\ref{app_cft}. This symmetry algebra is in fact enlarged to the contracted large ${\cal N}=4$ superconformal symmetry \cite{Maldacena:1999bp,Sevrin:1988ew}, however, we will not require the details of this in what follows. Exactly at the free point, this symmetry is boosted to include also a $\mathcal{W}_{\infty}$ algebra studied, for instance, in~\cite{Gaberdiel:2015mra,Gaberdiel:2015uca}.

Each ${\cal N} = 4$ algebra has an internal $\mathfrak{su}(2)$ R-symmetry, so there is an overall global symmetry $SU(2)_L\times SU(2)_R$. We denote the quantum numbers in these two $SU(2)$ factors as
\be
    SU(2)_L: ~(j, m)\qquad,\qquad SU(2)_R: ~ (\bar j, \bar m) \ .
\ee
In the geometrical setting of the CFT, this symmetry arises from the rotational symmetry of the $4$ spatial directions of $M_{4,1}$: we have $SO(4)_E \cong ( SU(2)_L\times SU(2)_R)/\mathbb{Z}_2$ and use spinor indices $\alpha, \bar\alpha$ for $SU(2)_L$ and $SU(2)_R$ respectively. Here the subscript $E$ stands for `external', which denotes that these rotations are in the noncompact directions. We also have another $SO(4)$ global symmetry in the four directions of the $T^4$. This symmetry we call $SO(4)_I$ (where $I$ stands for `internal'). This symmetry is broken by the compactification of the torus, but at the orbifold point it still provides a useful organising principle. We write $SO(4)_I\cong (SU(2)_1\times SU(2)_2)/\mathbb{Z}_2$ and use spinor indices $A, \dot A$ for $SU(2)_1$ and $SU(2)_2$ respectively.

\subsection{Neveu-Schwarz and Ramond sectors} \label{sec.sf}

The $\mathcal{N}=4$ superconformal algebra in two dimensions also has a non-trivial group of global automorphisms, with a particular $1$-parameter family, labelled by an angle $\pi\eta$, referred to as spectral flow\footnote{The full automorphism group of this $\mathcal{N}=4$ superconformal algebra is $SO(4)$ \cite{Schwimmer:1986mf}.}. These are maps between equivalent algebras and have the effect of changing the periodicity of the currents. In this section we very briefly review the rules for spectral flow transformations \cite{Schwimmer:1986mf} that will be used in the main computation of Section~\ref{sec.main}. Under spectral flow by $\eta$ units%
\footnote{We note that the left- and right-moving sectors of the $\mathcal{N}=(4,4)$ superconformal algebra can be spectral flowed independently by $\eta$ and $\bar{\eta}$ units respectively. In~\cite{Guo:2021uiu} a generalisation of this spectral flow was considered, where the generators of the free-field realisation of the $\mathcal{N}=4$ algebra can be decomposed into generators of two $\mathcal{N}=2$ algebras, which can be individually spectral flowed. This was dubbed `partial spectral flow' and will not be used here.}%
, the dimension $h$ and $J^3_0$ charge $m$ of an operator transform as
\begin{equation} \label{sfDims}
    h \rightarrow h' = h +m\eta +\frac{c \eta^2}{24} \quad\ ,\qquad m\rightarrow m' = m + \frac{c\eta}{12}\ ,
\end{equation}
where $c$ is the central charge%
\footnote{We note that the value of the central charge $c$ used in such spectral flow calculations depends on the state being considered. For a state on a single copy of the seed CFT one has $c=6$, whereas in general for a state on a $k$-wound component string one has $c=6k$. For a state of the full orbifold theory, the total central charge is $c=6N$.} of the CFT. Consider an operator $\mathcal{O}(w)$ with $J^3_0$ charge $m$. Under spectral flow by $\eta$ units at a point $w_0$ on the cylinder, this operator transforms as
\begin{equation} \label{sf_O}
    \mathcal{O}(w)\to \big(e^w-e^{w_0}\big)^{-\eta m}\mathcal{O}(w)\ ,
\end{equation}
if the corresponding state is annihilated by the mode $J^3_1$. In Appendix \ref{sec.SFcyl} we derive this condition on $\mathcal{O}(w)$ for the transformation rule \eqref{sf_O} to hold. In Appendix~\ref{sec.SFpl} it is found that a superconformal primary on the plane transforms similarly as
\begin{equation} \label{sf_01}
    \mathcal{O}(z)\to (z-z_0)^{-\eta m}\mathcal{O}(z)\ ,
\end{equation}
though the general condition on $\mathcal{O}(z)$ to have such a transformation is left for future work. For $\eta$ an odd integer, spectral flow interpolates between the NS- and R-sector boundary conditions for the fermionic degrees of freedom of the theory.

The superconformal primaries $\phi$ of the $\mathcal N=4$ superconformal algebra are defined, in the NS sector, by the mode conditions
\begin{equation} \label{def primary 2}
    L_{n}|\phi\rangle_{\scriptscriptstyle{N\!S}} = G^{\alpha}_{\dot A,r}|\phi\rangle_{\scriptscriptstyle{N\!S}} = J^{a}_{n}|\phi\rangle_{\scriptscriptstyle{N\!S}} =0 \quad,\quad n>0\ ,\ r\geq \frac12 \ .
\end{equation}
For a superconformal primary of the full theory, these symmetry current modes should be global modes -- for instance, the global Virasoro modes $L^{(g)}_{n}$ in the untwisted sector are given by the diagonal sum
\begin{equation} \label{globalL}
    L^{(g)}_{n} = \sum_{i=1}^N L^{(i)}_{n} \ ,
\end{equation}
where $L^{(i)}_{n}$ are the modes on the $i$th copy. A superconformal primary on a single copy would simply satisfy the conditions \eqref{def primary 2} with the modes on that particular copy. Under a spectral flow by $\eta=-1$ units to the R sector, the modes of the supercharges and $SU(2)$ currents transform as
\begin{equation} \label{flow super}
    G^{\pm}_{\dot A,r} \longrightarrow \,G^{\pm}_{\dot A,r\pm\frac{1}{2}} \quad,\quad J^{\pm}_n \longrightarrow J^{\pm}_{n\pm1} \ ,
\end{equation}
and so we see that the superconformal primary conditions in the NS sector \eqref{def primary 2} become the R-sector superconformal primary conditions
\begin{equation} \label{def primary 1}
\begin{aligned}
    L_{n}|\phi\rangle_R = G^{\alpha}_{\dot A,n}|\phi\rangle_R = J^{3}_{n}|\phi\rangle_R = J^{\pm}_{n\pm 1}|\phi\rangle_R =G^{-}_{\dot A,0}|\phi\rangle_R &=0 \quad,\quad n>0 \ .
\end{aligned}
\end{equation}
We note that for the NS and R sectors on the \emph{cylinder}, fermions have periodic boundary conditions in the R sector and anti-periodic boundary conditions in the NS sector. Mapping to the plane reverses these boundary conditions. These conditions dictate that fermions in the NS sector are half-integer modded and in the R sector are integer modded.

\subsection{Deformation away from the orbifold point}

The deformation of the CFT off the orbifold point is given by adding a deformation operator $D$ to the Lagrangian such that the action is modified as
\be\label{defor S}
    S\r S+\lambda \int d^2 z\, D(z, \bar z) \ ,
\ee
where $D$ has conformal dimensions $(h, \bar h)=(1,1)$ in order to be marginal. There are $20$ exactly marginal operators of the theory to choose from; $16$ of these correspond to $T^4$ shape and complex structure moduli, whilst the remaining $4$ are superdescendants of the twist-2 chiral/anti-chiral primaries $\sigma^{\alpha\bar\alpha}$ of the orbifold theory. The former set of moduli are in a sense `trivial deformations', whereas the latter four are `non-trivial' -- they break the higher-spin symmetry found at the orbifold point and move the theory towards the region with a semi-classical gravity description. A choice of $D$ that is a singlet under all of the $SU(2)$ symmetries at the orbifold point is
\be \label{D 1/4}
    D = \frac{1}{4}\epsilon^{\dot A\dot B}\epsilon_{\alpha\beta}\epsilon_{\bar\alpha \bar\beta} \,G^{\alpha}_{\dot A, -\h} \bar G^{\bar \alpha}_{\dot B, -\h} \sigma^{\beta \bar\beta} \ .
\ee
Here $G$ and $\bar G$ are respectively the left- and right-moving supercharge modes at the \emph{orbifold point}, \textit{i.e.} at $\lambda=0$. This choice of deformation operator is the projection onto the singlet representation of $SU(2)_2$; the remaining three non-trivial deformation operators form the triplet projection. The $SO(4)_I\,$-invariant operator \eqref{D 1/4} corresponds to turning on the dual string tension. The operator $\sigma^{\alpha \bar\alpha}$ is made by adding R-charge to the bare twist operator $\sigma_2$ to form a chiral/anti-chiral primary. In the particular case of $k=2$, this is done using only spin fields \cite{Lunin:2001fv} to give
\begin{equation} \label{eq.sig2}
    \sigma^{\alpha\bar{\alpha}} = S_2^{\alpha} \bar{S}_2^{\bar{\alpha}} \sigma_2 \ .
\end{equation}
These spin fields change the fermion boundary conditions around their insertion points. The operator $\sigma_2$ twists together two copies, $i$ and $j$, of the $c=6$ seed CFT, which in an $S_N$-invariant form is written as
\begin{equation}
    \sigma_2 = \sum_{\substack{i,j=1\\j<\:\!i}}^N \sigma_{(ij)} \ .
\end{equation}

\subsection{The Ramond ground states} \label{sec R ground}

Consider the Ramond sector of the theory. On a $k$-twisted component string, the ground state is degenerate in both the left and right sectors. There are $4$ Ramond ground states for the left sector per value of $k$, which we denote as
\be \label{qgroundstates}
    |0^-\rangle_{R}^{[k]}\ \ ,\ \ |0^+\rangle_R^{[k]}=d^{++[k]}_0d^{+-[k]}_0|0^-\rangle_R^{[k]}\ \ ,\ \ |0\rangle_R^{[k]}=d^{++[k]}_0|0^-\rangle_R^{[k]}\ \ ,\ \ |\tilde 0\rangle_R^{[k]}=d^{+-[k]}_0|0^-\rangle_R^{[k]} \ ,
\ee
where the $d^{\alpha A}_n$ are the Ramond-sector modes of the canonical free fermions of the orbifold CFT. Our conventions for the anti-commutator for these fermion modes is given in \eqref{eq.ddcomm}. A similar set of $4$ Ramond ground states exist for the right sector. The superscript $[k]$ denotes that these states are in the $k$-twisted sector of the theory. Thus for any component string (\textit{i.e.} for each twisted sector), there are $16$ Ramond-Ramond ground states. We will label these ground states as $|L,R\rangle_R^{[k]}$ where $L=0^-,0^+,0,\tilde 0$ and $R=\bar0^-,\bar0^+,\bar0,\bar{\tilde{0}}$. The Ramond-Ramond ground state dimensions are $h=\bar{h}=\tfrac{c}{24}$ and with the conventions of Section~\ref{sec.sf}, they are related by a spectral flow of $\eta=+1$ to chiral primaries in the NS sector. The Ramond ground state $|0^-\rangle_R^{[1]}$ is then the spectral flow of the NS vacuum $|0\rangle_{\scriptscriptstyle{\!N\!S}}$, having $h=m=0$. Whilst we use only the first of the Ramond ground states given in \eqref{qgroundstates} for the particular lifting computation described in this paper, we hope to return to the more general problem for arbitrary ground states in future work.

\subsection{Lift formulae} \label{sec:LiftForm}

Using conformal perturbation theory around the symmetric orbifold theory with the deformation \eqref{defor S}, one finds that the change in total energy of a state to second order in $\lambda$ involves the evaluation of a four-point correlation function containing two twist operators of the form \eqref{D 1/4} (see, for instance, \cite{Hampton:2018ygz}). Seemingly the most straightforward method is to just calculate this four-point function and integrate over the insertion points of the deformation operators. Correlation functions of twist operators have been considered variously in \cite{Arutyunov:1997gi,Arutyunov:1997gt,Lunin:2000yv,Lunin:2001fv,Pakman:2009ab,Pakman:2009mi,Pakman:2009zz,Burrington:2012yn,Burrington:2012yq,Roumpedakis:2018tdb,GarciaiTormo:2018vqv,Burrington:2018upk,Dei:2019iym} and this method has been used for lifting computations of simple states in \cite{Gaberdiel:2015mra,Hampton:2018ygz}. This method requires the use of a covering space to resolve the effect of the twist operator insertions. Whilst the covering space can be topologically a sphere, in which case correlation functions are usually tractable, it is also possible for it to be of higher genus (seen from the Riemann-Hurwitz theorem). In this case the lifting problem is far more difficult. 

Alternatively, the method of Gava-Narain \cite{Gava:2002xb} proposed a way to compute the lift of right-chiral states (also referred to as D1-D5-P states) in terms of a finite sum of 3-point functions on the plane, involving the twist operator \eqref{eq.sig2} and the unperturbed state. This method was elucidated from the Hamiltonian (cylinder) perspective in \cite{Guo:2019pzk}. The Gava-Narain method was used to compute lifts in \cite{Guo:2019ady,Guo:2020gxm,Benjamin:2021zkn}. Since only 3-point functions are required and crucially only \emph{one} twist operator, any covering space will be genus $0$ and so is much simpler than the general 4-point function case described above. However, despite requiring a finite number of 3-point functions, this can still be a large enough number to make the first method simpler in some cases.

In this paper, we will opt for a so-called `hybrid method' in which the straight calculation of 4-point functions is used as in the first method above, but then instead of performing integrals over the deformation operator insertions, we implement a projection of intermediate states onto the level-$n$ subspace~\cite{Gaberdiel:2015uca,Benjamin:2021zkn}. We find that with the particular choice of states considered in what follows, this method is the most efficient and affords us the desired level of generality.

Consider the subspace formed by unperturbed states (\textit{i.e.} states at the free point) at level-$n$ in the R sector, which we denote as $\{\mathcal{O}^{(0)}_{a}\}$ with dimensions $(h,\bar h)=(n+\frac{c}{24},\frac{c}{24})$. These states have only left-moving excitations above the Ramond-Ramond ground states. We will denote the projector onto this subspace as $\mathcal{P}$. As we move away from the orbifold point of the moduli space, the dimensions of the D1-D5-P states can change
\be
    (h,\bar h)\to (h+\delta h, \bar h + \delta \bar h) \ .
\ee
Since the conformal spin $|h-\bar h|$ must be an integer, it implies that the left- and right-moving anomalous dimensions must be equal: \textit{i.e.} that 
\be
    \delta h = \delta \bar h \ .
\ee
Furthermore, because Ramond ground states have the lowest possible dimension in that sector, the change in dimension for the unexcited right-moving part of D1-D5-P states must be non-negative, $\delta \bar h\geq 0$. Therefore, the total energy lift must be non-negative
\be \label{E p}
E^{(2)}=2 \delta h \geq 0 \ .
\ee
This non-negativity of the lift implies that the first nontrivial contribution starts at second order in the perturbation $\lambda$, so $\delta h \sim O(\lambda^2)$.

Within the level-$n$ subspace, define the lifting matrix
\begin{equation} \label{liftmatrix 1}
    E^{(2)}_{ba} = 2 \lambda^2 \Big\langle \mathcal{O}^{(0)}_{b}\Big| \Big\{ \bar G^{+(P)\dagger}_{+,0}, \bar G^{+(P)}_{+,0} \Big\} \Big|\mathcal{O}^{(0)}_{a}\Big\rangle =2 \lambda^2 \Big\langle \mathcal{O}^{(0)}_{b}\Big| \Big\{ \bar G^{+(P)\dagger}_{-,0}, \bar G^{+(P)}_{-,0} \Big\} \Big|\mathcal{O}^{(0)}_{a}\Big\rangle \ ,
\end{equation}
with the $\bar G^{\bar \alpha (P)}_{\dot A,0}$ operators being defined as
\be\label{GN p s}
    \bar G^{\bar \alpha (P)}_{\dot A,0} \equiv \pi \mathcal P G^{+}_{\dot A,-\frac{1}{2}}\sigma^{-\bar \alpha} \mathcal P = - \pi \mathcal P G^{-}_{\dot A,-\frac{1}{2}}\sigma^{+\bar \alpha} \mathcal P \ ,
\ee
where we note that there is some redundancy in the projection operator when we insert (\ref{GN p s}) into (\ref{liftmatrix 1}). Our conventions for Hermitian conjugation can be found in Appendix~\ref{app.Herm}, with the conjugation of the $\bar G^{\bar \alpha (P)}_{\dot A,0}$ being Equation~\eqref{conju}. Due to the projection operator $\mathcal{P}$, \eqref{GN p s} does not depend on the position of the operator of the form $G\sigma$. This can be seen by considering the Hamiltonian evolution on the Euclidean cylinder $(\tau,\sigma)$ of an initial state with dimension $h$ from $\tau=\tau_i$ to $\tau=\tau'$, at which an operator is inserted (the two deformation operators in our case). This insertion changes the state to one of dimension $h'$ which then evolves up to some $\tau=\tau_f$. In total this evolution gives a factor of $\sim e^{-h(\tau'-\tau_i)} e^{-h'(\tau_f-\tau')}$ and so if one demands that the state created at $\tau'$ has $h'=h$ then the evolution factor becomes independent of $\tau'$. It has been shown in \cite{Guo:2019pzk} that $\bar G^{\bar \alpha (P)}_{\dot A,0}$ is related to the first order (in $\lambda$) correction to the supercharge $\bar G^{\bar \alpha}_{\dot A,0}$ and that the second-order lift does not depend on the first-order correction to the state. The eigenvalues of the lifting matrix \eqref{liftmatrix 1} are the values of the second order lift for the various operators at a fixed level. The corresponding eigenstate is the zeroth order state having this lift.

In the computation that follows, we will be interested solely in the average lift for a level-$n$ state (of the zeroth order theory) $\Phi\in\{\mathcal{O}^{(0)}_{a}\}$, given as the weighted sum\vspace{4pt}
\begin{equation}
    E^{(2)}(\Phi) = \sum_{a'} \big| \langle \widetilde{\mathcal{O}}_{a'}| \Phi\rangle\big|^2 E^{(2)}_{a'} \ ,\vspace{-5pt}
\end{equation}
with $\{\widetilde{\mathcal{O}}_{a'}\}$ being the set of good eigenstates of the $O(\lambda^2)$ Hamiltonian, with lift $E^{(2)}_{a'}$. The set of $O(\lambda^0)$ level-$n$ energy eigenstates $\{\mathcal{O}_{a}\}$ are mixed together into the set of energy eigenstates $\{\widetilde{\mathcal{O}}_{a'}\}$ of the $O(\lambda^2)$ theory. Because eigenvalues of $E^{(2)}_{ba}$ are non-negative, if the above average is positive, the state $\Phi$ must contain a lifted state. Expanding the anti-commutator of \eqref{liftmatrix 1} and using the conjugation relations \eqref{conju}, this lifting can be written as
\begin{equation} \label{lift average}
     E^{(2)}(\Phi) = 2\lambda^2\pi^2\langle \Phi\big|\Big[( G^{+}_{-,-\h}\sigma^{--})\mathcal{P}( G^{-}_{+,-\h}\sigma^{++}) + ( G^{-}_{+,-\h}\sigma^{++})\mathcal{P}(G^{+}_{-,-\h} \sigma^{--}) \Big]\big|\Phi\rangle \ .
\end{equation}
This formula was derived in \cite{Hampton:2018ygz} and calculating the lift in this way is what we refer to as the hybrid method.

\section{Lifting of single-copy superconformal primaries} \label{sec.main}

\subsection{The states} \label{sec.states}

The Ramond-sector states under consideration in this paper are in the $k=1$, untwisted sector of the orbifold theory (states of $N$ singly-wound strings) where one copy (say copy 1) is excited into a left-moving superconformal primary state $|\phi\rangle_R^{[1](1)}$, with the right-moving part being the Ramond ground state $|\bar0^-\rangle_R^{[1](1)}$. This single-copy superconformal primary satisfies the conditions \eqref{def primary 1} for the single-copy modes and has dimension $h_R$ and charge $m_R$. The remaining \mbox{$N-1$} copies are in the Ramond-Ramond ground state $|0^-,\bar0^-\rangle_R^{[1](j)}$. To avoid overly cumbersome notation, we will drop the twist-sector label $[k]$ since we consider only the singly-twisted case in the present paper. Hence, the class of states we consider is of the form
\be \label{eq.state1}
    |\Phi\rangle^{(1)}_{R} \equiv |\phi,\bar0^-\rangle_R^{(1)} |0^-,\bar0^-\rangle_R^{(2)} \cdots |0^-,\bar0^-\rangle_R^{(N)} \ .
\ee
In order for this to be a physical state of the orbifold theory, it is necessary to then symmetrise over the choice of excited copy; with an appropriate normalisation factor this gives the state
\begin{equation} \label{FullState}
    |\Phi\rangle \equiv \frac1{\sqrt{N}} \sum_{i=1}^{N} \bigotimes_{j\neq i}^{N} |\phi,\bar0^-\rangle_R^{(i)} |0^-,\bar0^-\rangle_R^{(j)} = \frac1{\sqrt{N}} \sum_{i=1}^{N} |\Phi\rangle^{(i)}_{R} \ .
\end{equation}
Due to the twist operators in \eqref{lift average} acting on pairs of copies and the fact that we are computing an expectation value, this forces the second twist operator to act on the same pair as the first in order to undo the twist. Therefore, at any intermediate step of the computation it is sufficient to work with only a particular ordered pair of copies, with the final lifting for generic values of $N$ gaining an extra combinatoric factor. With the initial and final states being of the form \eqref{FullState}, there are two distinct cases possible when choosing two copies to twist together. Either both copies are in the Ramond-Ramond ground state $|0^-,\bar0^-\rangle_R$ or one of the copies is in the superconformal primary state $|\phi,\bar0^-\rangle_R$. The former case would yield an expectation value that computes the lift of $|0^-,\bar0^-\rangle_R\otimes|0^-,\bar0^-\rangle_R$, which vanishes since all Ramond ground states are unlifted, leaving terms only of the latter type. Since we have only singly-wound strings, the first twist operator can only twist together two of these to form a doubly-wound component string (and cannot break apart component strings). This ensures that the covering space necessary to resolve the twist operators in the correlation function \eqref{lift average} will have genus $0$.

A further simplification occurs by considering the right-moving part of the second term in \eqref{lift average}. In \eqref{FullState} every copy is in a right-moving Ramond ground state and so, focusing on the particular term \eqref{eq.state1}, we can use the right-moving relations (derived in \cite{Guo:2019ady,Guo:2020gxm})
\begin{subequations} \label{sigsig}
\begin{align}
    \Robmb{2}\Robmb{1}\, \bar{\sigma}^{-}\, \mathcal{P}\, \bar{\sigma}^{+} \Robmk{1}\Robmk{2} &= 1 \ , \label{sigsig1}\\
    \vspace{3pt} \Robmb{2}\Robmb{1}\, \bar{\sigma}^{+} \,\mathcal{P}\, \bar{\sigma}^{-} \Robmk{1}\Robmk{2} &= 0 \label{sigsig2} \ ,
\end{align}
\end{subequations}
where we have dropped the $N-2$ vacuum copies not participating in the twisting. Thus, we get a particular left-moving two-copy contribution to the lift \eqref{lift average} of%
\footnote{It turns out that the twist operators $\sigma^{\alpha\bar{\alpha}}$ cannot be consistently split into left- and right-moving parts. As explained in Appendix~\ref{app.Herm}, it is still possible to define a set of notational conventions allowing us to use the splitting $\sigma^{\alpha\bar{\alpha}}= \sigma^{\alpha}\bar{\sigma}^{\bar{\alpha}}$, given that an additional minus sign is included once the left- and right-moving parts of our computation are brought back together in \eqref{LiftLeft2}. This allows for drastically more convenient written expressions.}%
\begin{equation} \label{LiftLeft2}
     E^{(2)}_{(1)(1)}(\Phi) \equiv -2\lambda^2\pi^2 \Romb{2}\phiRb{1} (G^{+}_{-,-\h}\sigma^{-})\mathcal{P}( G^{-}_{+,-\h}\sigma^{+}) \phiRk{1}\Romk{2} \ ,
\end{equation}
where the subscript $(1)(1)$ on the lift refers to the excited copy in the initial and final states.

We now discuss the combinatoric factor necessary for obtaining the general $N$ result from \eqref{LiftLeft2}. For the untwisted-sector state we consider, in which only one copy is excited, there will be a nonzero expectation value only when the first twist operator acts on a pair of copies that includes an excited copy. The second twist operator then has to act on this same pair in order to untwist them. The `diagonal' terms are expectation values where the final state has the same copy excited as in the initial state. There are $N$ different copies that could be excited and then $N-1$ choices of the second, vacuum, copy. Thus there are $N(N-1)$ diagonal terms. In the notation of \eqref{LiftLeft2}, these would be of the form $E^{(2)}_{(i)(i)}$ for some $1\leq i\leq N$. The `off-diagonal' terms have, for a given choice of excited copy in the initial state ($N$ choices), a different copy excited in the final state ($N-1$ such choices). These choices of excited states then force the respective choice of vacuum copy in the initial and final states. Thus there are $N(N-1)$ off-diagonal terms. In the notation of \eqref{LiftLeft2}, these would be of the form $E^{(2)}_{(j)(i)}$ for $i\neq j$. In both cases the remaining $N-2$ vacuum copies are spectator copies, playing no role in the lifting calculation. Along with the normalisation factor in \eqref{FullState} (for the initial and final states), the total lift is then given in terms of \eqref{LiftLeft2} as
\begin{equation} \label{LiftLeftN}
    E^{(2)}_{h,m}(\Phi) = \bigg(\frac1{\sqrt{N}}\bigg)^{\!2}\, \Big(N(N-1)+N(N-1)\Big) E^{(2)}_{(1)(1)}(\Phi) = 2(N-1) E^{(2)}_{(1)(1)}(\Phi)\ .
\end{equation}
We will proceed by computing the contribution to the lift from the particular diagonal term \eqref{LiftLeft2} and then will use \eqref{LiftLeftN} to get the full lift. We note that while permutation invariance dictates that each of the diagonal terms $E^{(2)}_{(i)(i)}$ will contribute the same amount to the lift, and likewise for each of the $E^{(2)}_{(j)(i\neq j)}$, at this stage it is not at all clear that a diagonal and an off-diagonal term will contribute equally. This will be addressed at the end of the computation of $E^{(2)}_{(1)(1)}$ in Section~\ref{LiftComp}, with its equality to $E^{(2)}_{(2)(1)}$ explicitly shown in Appendix~\ref{app.B}.

\subsubsection{Single-copy superconformal primaries} \label{sec:SingleSCP}

We now give some comments on single-copy superconformal primary states that will significantly narrow down the possible values of dimension and charge that need to be considered. Our comments will be in the language of the NS sector, but these can all be mapped to the Ramond sector. Initially it appears that we can have arbitrary left-moving dimension $h$ and $J^3_0$ charge $m$ for the untwisted-sector state where one copy is excited to a left-moving superconformal primary and the remaining $N-1$ copies being the NS vacuum. However, this turns out not to be the case. Consider the norm
\begin{equation}
    \big|\big| J^+_{-n}|\phi\rangle\big|\big|^2 \geq 0 \ ,
\end{equation}
for $n>0$, which is positive definite in a unitary theory. From the commutator \eqref{JJcomm} and the definition $J^{\pm}_{n}\equiv J^1_n \pm iJ^2_n$, we find that for a superconformal primary
\begin{align} \label{mConstraint1}
    0\leq \langle\phi| J^-_{n}J^+_{-n} |\phi\rangle = \langle \phi| (Nn-2J^3_0)|\phi\rangle \ ,
\end{align}
where we have used the conditions \eqref{def primary 2}. For $|\phi\rangle$ being a state on a single copy (for which $N=1$) with a $J^3_0$ charge of $m$, \eqref{mConstraint1} then gives the set of constraints (one per value of $n$)
\begin{equation} \label{mConstraint2}
    m \leq \frac{n}{2} \ .
\end{equation}
Specifically, the most stringent constraint comes from $n=1$ for which we find that $m\leq \tfrac12$. Since the action of the modes $J^{\pm}_0$, which act as raising and lowering operators for the eigenvalue of $J^3_0$ within an $SU(2)_L$ multiplet, commutes with the superconformal primary conditions \eqref{def primary 2} the constraint \eqref{mConstraint2} must be satisfied by all members of an $SU(2)_L$ multiplet. Labelling such multiplets by the eigenvalue of the $SU(2)_L$ quadratic Casimir $j$, since the range of $m$ within a given multiplet is $m=-j,-j+1,\dots, j-1, j$ we see that there are only two possibilities. Single-copy superconformal primaries are either in $j=\tfrac12$ or $j=0$ multiplets.

From \eqref{mConstraint1}, for a single-copy superconformal primary with $m=\tfrac12$ we find
\begin{equation}
    \langle\phi|J^-_1J^+_{-1}|\phi\rangle =0 \quad\Longrightarrow\quad J^+_{-1}|\phi\rangle =0 \ .
\end{equation}
The state $|\phi\rangle$ thus satisfies
\begin{equation}
    G^+_{\dot{A},-\frac12}|\phi\rangle = \Big[J^+_{-1},G^-_{\dot{A},\frac12}\Big]|\phi\rangle = 0 \ ,
\end{equation}
which is exactly the condition for $|\phi\rangle$ to be chiral. Since chiral primaries are unlifted, this leaves us simply with $j=0$ multiplets. It should be noted that this argument only holds true for superconformal primaries on a single copy, \textit{i.e.} when $c=6$. For instance, in higher twist sectors it is possible to have superconformal primaries with higher values of $j$.

\subsection{Lift computation} \label{LiftComp}

To evaluate \eqref{LiftLeft2}, we will instead compute the `left-moving' cylinder amplitude
\be \label{amp}
    A^{(1)(1)}(w_2,w_1) \equiv \Romb{2} \phiRb{1} (G^{+}_{-,-\h}\sigma^{-})(w_2)\, (G^{-}_{+,-\h}\sigma^{+})(w_1) \phiRk{1}\Romk{2} \ ,
\ee
and then project onto the term independent of $e^{w_2-w_1}$ in order to enact the projection $\mathcal{P}$. The first step in the computation of this amplitude is to replace the modes $G^{\alpha}_{\dot A,-\h}$ by contour integrals of the associated operators $G^{\alpha}_{\dot A}(w)$ to produce a sum of amplitudes with the $G$ modes acting in various combinations on the initial and final states.

\subsubsection*{Unwrapping the supercharge contours}

Using the standard relation
\be
    \big(G^{\alpha}_{\dot A,-\h}\sigma^{\beta}\big)(w_2) =  \oint_{C_{w_2}} \frac{dw}{2\pi i}\,G^{\alpha}_{\dot A}(w) \sigma^{\beta}(w_2) \ ,
\ee
where $C_{w_2}$ is a contour centred on the point $w_2$, the contours of the two supercharge modes in \eqref{amp} can be `unwrapped'. Starting with the contour of the $G^+_-$ supercharge, this can be unwrapped to give%
\footnote{We note that the supercharge modes around the two twist operators are written with half-integer modes, despite stating that we are working in the R sector on the cylinder. This is because the even twist operators contain spin fields, as in \eqref{eq.sig2}, which change the fermion boundary conditions around those specific points from R to NS. Once these mode contours are unwrapped, acting on the initial and finial states, they are integer moded.}%
\begin{align} \label{unwrap1}
    A^{(1)(1)}(w_2,w_1) =&\ \Romb{2}\phiRb{1} (G^{+}_{-,0})\, \sigma^{-}(w_2)\, \big(G^{-}_{+,-\h}\sigma^{+}\big)(w_1)\,\phiRk{1}\Romk{2}\nn
    &\, - \Romb{2}\phiRb{1}\, \sigma^{-}(w_2)\, \big(G^{+}_{-,-\h}G^{-}_{+,-\h}\sigma^{+}\big)(w_1)\,\phiRk{1}\Romk{2}\nn
    &\, + \Romb{2}\phiRb{1}\, \sigma^{-}(w_2)\, \big(G^{-}_{+,-\h}\sigma^{+}\big)(w_1)\,(G^{+}_{-,0})\phiRk{1}\Romk{2} \ .
\end{align}
In the second term of \eqref{unwrap1} the minus sign is due to the relative directions of the unwrapped $G^{+}_{-}$ contour and the existing $G^-_+$ contour, whilst the third term also has one sign from reversing the contour direction and one from moving the $G^+_-$ through the $G^-_+$. The first term of \eqref{unwrap1} vanishes using \eqref{def primary 1} due to $G^+_-$ annihilating the final-state superconformal primary. Using the anti-commutator \eqref{GGcomm} and the relations in \eqref{A3}, the operator at $w_1$ in the second term of \eqref{unwrap1} becomes
\begin{equation}
    G^{+}_{-,-\h}G^{-}_{+,-\h}\sigma^{+} = \big\{G^{+}_{-,-\h},G^{-}_{+,-\h}\big\}\,\sigma^{+} = \p \sigma^{+} \ ,
\end{equation}
and we obtain the amplitude
\begin{align} \label{unwrap11}
    A^{(1)(1)}(w_2,w_1) &= - \Romb{2}\phiRb{1}\, \sigma^{-}(w_2)\, \p\sigma^{+}(w_1)\,\phiRk{1}\Romk{2} \nn
    &\quad\,+ \Romb{2}\phiRb{1}\, \sigma^{-}(w_2)\, \big(G^{-}_{+,-\h}\sigma^{+}\big)(w_1)\,(G^{+}_{-,0})\phiRk{1}\Romk{2} \ .
\end{align}
We now continue by unwrapping the $G^-_+$ contour in the second term of \eqref{unwrap11}. Due to $G^-_+$ annihilating the anti-chiral primary $\sigma^-$, this unwrapping generates only two nonzero terms and the amplitude becomes
\begin{align} \label{amp 1}
    A^{(1)(1)}(w_2,w_1) &= - \Romb{2}\phiRb{1}\, \sigma^{-}(w_2)\, \p\sigma^{+}(w_1)\,\phiRk{1}\Romk{2} \nn
    &\quad\, +\Romb{2}\phiRb{1}(G^{-}_{+,0})\, \sigma^{-}(w_2)\, \sigma^{+}(w_1)\,(G^{+}_{-,0})\phiRk{1}\Romk{2} \nn
    &\quad\, -\Romb{2}\phiRb{1}\,\sigma^{-}(w_2)\, \sigma^{+}(w_1)\,(G^{-}_{+,0}G^{+}_{-,0})\phiRk{1}\Romk{2} \ .
\end{align}

\subsubsection*{Spectral flowing to the NS sector}

Now that the amplitude \eqref{amp} has been converted to a sum of simpler amplitudes containing only twist operators without surrounding $G$ contours, the next step is to spectral flow to the NS sector. We do this because the Ramond ground state $|0^-\rangle_R$ can be flowed to the unique NS vacuum $|0\rangle_{\scriptscriptstyle{N\!S}}$ with $h=j=0$ and so will simplify the eventual calculation of the amplitude on the covering space. We note that one can choose to first spectral flow the amplitude \eqref{amp} to the NS sector directly, before unwrapping contours. However, in that case the supercharges become anti-periodic on the cylinder and their contours must be unwrapped with more care.

Consider the component amplitude
\begin{equation} \label{CompAmp}
    \widetilde{A}(w_2,w_1) \equiv \Romb{2}\phiRb{1}\, \sigma^{-}(w_2)\, \sigma^{+}(w_1)\,\phiRk{1}\Romk{2} \ .
\end{equation}
This amplitude will be used to compute each of the terms in \eqref{amp 1}. Under a spectral flow by $\eta=1$ units around $\tau=-\infty$, the modes of the supercharges transform as in \eqref{flow super}, from which we see that the superconformal primary conditions in the R sector \eqref{def primary 1} become those of the NS sector \eqref{def primary 2}. Therefore, our state $|\phi\rangle_R$ spectral flows to an NS-sector superconformal primary $|\phi\rangle_{\scriptscriptstyle{N\!S}}$. The quantum numbers of this NS-sector state we denote as $h$ and $m$ for the dimension and charge respectively. As explained in Section~\ref{sec:SingleSCP}, single-copy superconformal primary states can only be in $j=\tfrac12$ multiplets, in which case they are chiral and unlifted, or in $j=0$ multiplets. We therefore have $m=0$ for the state $|\phi\rangle_{\scriptscriptstyle{N\!S}}$, although sometimes we keep $m$ general for later convenience. For the twist operators in \eqref{CompAmp}, using \eqref{sf_O} these transform under this spectral flow as
\begin{equation}
    \sigma^+(w_1) \longrightarrow e^{-\frac{1}{2}w_1}\sigma^+(w_1) \quad,\quad \sigma^-(w_2) \longrightarrow e^{\frac{1}{2}w_2}\sigma^-(w_2) \ .
\end{equation}
Since the amplitude \eqref{CompAmp} is invariant under spectral flow\footnote{A result that is also important here is that also the lift for D1-D5-P states \eqref{lift average} is invariant under spectral flow and not just the amplitude \eqref{amp}. In other words, the value of the lift for states related by spectral flow is the same. In order to see this, it is necessary to check how the projection operator $\mathcal{P}$ transforms. This can be found in \cite{Guo:2021uiu}.}, we write
\begin{equation}
    \widetilde{A}(w_2,w_1) = \widetilde{\mathcal{A}}(w_2,w_1;\phiNS) \, e^{\frac{1}{2}(w_2-w_1)}\ ,
\end{equation}
where
\begin{equation} \label{Atilde}
    \widetilde{\mathcal{A}}(w_2,w_1;\phiNS) \equiv \NSob{2}\!\!\phiNSb{1}\, \sigma^{-}(w_2) \,\sigma^{+}(w_1)\, \phiNSk{1}\,\NSok{2} \ .
\end{equation}
By using the transformations \eqref{flow super}, the anti-commutator \eqref{GGcomm} and the fact that $G^{-}_{+,\h}$ annihilates the initial state, we can compute the amplitude \eqref{amp 1} from the NS-sector expression
\begin{equation} \label{amp 2}
\begin{aligned}
    A^{(1)(1)}(w_2,w_1) = -\partial_{w_1}\Big(\widetilde{\mathcal{A}}(w_2,w_1;\phiNS) e^{\frac{1}{2}(w_2-w_1)}\Big) &+ \widetilde{\mathcal{A}}\big(w_2,w_1;G^{+}_{-,-\h}\phiNS\big) \, e^{\frac{1}{2}(w_2-w_1)} \\
    & -h\,\widetilde{\mathcal{A}}(w_2,w_1;\phiNS)\, e^{\frac{1}{2}(w_2-w_1)} \ .
\end{aligned}
\end{equation}

\subsubsection*{The base amplitude}

In \cite{Hampton:2018ygz}, the base vacuum amplitude
\begin{equation} \label{BaseAmp}
    \widetilde{\mathcal{A}}(w_2,w_1;\NSo) = \NSob{2}\!\!\NSob{1}\, \sigma^{-}(w_2) \,\sigma^{+}(w_1)\, \NSok{1}\,\NSok{2} \ ,
\end{equation}
was computed. The key steps in that computation are as follows:
\begin{enumerate}[start=1,
    labelindent=\parindent,
    leftmargin =1.1\parindent,
    label=(\arabic*)]
\item \label{base1}\vspace{5pt}
The cylinder amplitude $\widetilde{\mathcal{A}}(w_2,w_1;\NSo)$ is first mapped to the plane via the conformal transformation
    \begin{equation} \label{wtoz}
        w\to z(w) = e^{w} \ .
    \end{equation}
    
\item \label{base2}
On the plane $(z,\bar z)$, local operators are not single-valued due to the presence of twist operators in our amplitudes. These twist operators induce non-trivial monodromies around their insertion points. By mapping to the covering space, this effect of the twist operators can be removed. The covering map $z\to t$ in this case is of the form
\begin{equation} \label{CoverMap}
    z(t) = \frac{(t+a)(t+b)}{t} \ ,
\end{equation}
with the locations of the twist operators being the branch points
\begin{equation} \label{t1t2}
    t_1 = -\sqrt{ab} \ \ ,\quad t_2=\sqrt{ab} \quad\text{s.t.}\quad \frac{dz}{dt} = \frac{(t-t_1)(t-t_2)}{t^2} \ .
\end{equation}
Clearly inverting the map \eqref{CoverMap} yields two branches, corresponding to the two (ramified) sheets of the covering surface. The covering space in this particular case has genus $0$.

\item \label{base3}
For the case of even twisted sectors, in this case $k=2$, the twist operators $\sigma_2$ leave a spin field $S_2$ at their insertion points in the covering space -- see \eqref{eq.sig2}. These spin fields change the fermion boundary conditions around their insertion points, in this case from NS to R, and thus can be removed by appropriate spectral flows. We spectral flow by $\eta = -1$ units around the point $t=t_1$ to remove the spin field from $\sigma^{+}$ and by $\eta = 1$ units around the point $t=t_2$ to remove the spin field from $\sigma^{-}$.

\item \label{base4}
In the case of the vacuum base amplitude \eqref{BaseAmp}, there are now no operator insertions left on the covering space and the remaining amplitude is simply
\begin{equation}
    \NSob{2}\!\!\NSob{1} 0\rangle^{(1)}_{\scriptscriptstyle{N\!S}}\NSok{2}=1 \ .
\end{equation}
\end{enumerate}\vspace{5pt}
The resulting vacuum amplitude is given by \cite{Hampton:2018ygz}
\begin{equation} \label{BaseAmpResult}
    \widetilde{\mathcal{A}}(w_2,w_1;\NSo) = \frac{a-b}{4\sqrt{ab}} \ ,
\end{equation}
where $a$ and $b$ are given in terms of the insertion points on the cylinder by
\begin{equation} \label{abDef}
    a = e^s \cosh^2\!\Big(\frac{\Delta{w}}4\Big) \quad,\quad b = e^s \sinh^2\!\Big(\frac{\Delta{w}}4\Big) \ ,
\end{equation}
where we define
\begin{equation} \label{swDef}
    s \equiv \h(w_1+w_2) \quad,\quad \Delta w\equiv w_2-w_1 \ .
\end{equation}

\subsubsection*{The component amplitude}

The component amplitude \eqref{Atilde} necessary for the lifting computation in the present paper can now be found in terms of the base amplitude \eqref{BaseAmpResult}. Our amplitude will simply contain extra factors due to the transformation of the initial- and final-state superconformal primaries under each of the steps \ref{base1}-\ref{base4} above. The corresponding factors are:
\begin{enumerate}[start=1,
    labelindent=\parindent,
    leftmargin =1.1\parindent,
    label=(\roman*)]
\item \label{comp1} \vspace{5pt}
Under a conformal map $z\to z'$, a conformal primary $\mathcal{O}_{h,j}$ transforms as
\begin{equation} \label{PrimTrans}
    \mathcal{O}_{h,j}(z) \longrightarrow \bigg(\frac{dz'}{dz}\bigg)^{\!h}\, \mathcal{O}_{h,j}(z') \ .
\end{equation}
From the map to the plane \eqref{wtoz}, the twist operators pick up factors of $z_1^{1/2}$ and $z_2^{1/2}$ which are contained within the result \eqref{BaseAmpResult} and the initial and final states in \eqref{Atilde} do not generate additional non-trivial factors.

\item \label{comp2}
Under the map to the covering space \eqref{CoverMap}, the initial state on the first copy transforms, using \eqref{PrimTrans}, as
\begin{equation} \label{iCoverFactor}
    \phiNS\big(z=0,(1)\big)\longrightarrow \left(\frac{dt}{dz}\right)^{\!h} \phiNS(t=-a) = \left(\frac{a}{a-b}\right)^{\!h} \phiNS(t=-a) \ ,
\end{equation}
where \eqref{t1t2} has been used on the right-hand side. The point $z=\infty$, which is the location of the final state on copy 1, maps to $t=\infty$ in the covering space. Since $z\sim t$ as $t\rightarrow \infty$, we find
\be \label{fCoverFactor}
    \phiNS\big(z\rightarrow \infty,(1)\big) \longrightarrow \phiNS(t\rightarrow \infty) \ .
\ee

\item \label{comp3}

Although $|\phi\rangle_{\scriptscriptstyle{N\!S}}$ has $m=0$ and so the initial and final states transform trivially under the spectral flows from step \ref{base3} above, we keep $m$ general here for later convenience. Therefore, using \eqref{sf_O} and the results of Appendix~\ref{app.SF} we find that the initial state in the covering space transforms as
\begin{equation}
    \phiNS(t=-a) \longrightarrow (-a-t_1)^m (-a-t_2)^{-m}\, \phiNS(t=-a) = \left(\frac{\sqrt{a}-\sqrt{b}}{\sqrt{a}+\sqrt{b}}\right)^{\!m} \phiNS(t=-a) \,.
\end{equation}
Equivalently, for the final state we find the transformation
\begin{equation}
    \phiNS(t\rightarrow \infty) \longrightarrow \phiNS(t\rightarrow \infty) \ .
\end{equation}

\item \label{comp4}
For the amplitude $\widetilde{\mathcal{A}}(w_2,w_1;\phiNS)$, the equivalent of step \ref{base4} above leaves us instead with
\begin{equation}
    \NSob{2}\!\!\phiNSb{1} \phi\rangle^{(1)}_{\scriptscriptstyle{N\!S}}\NSok{2}=1 \ ,
\end{equation}
as the remaining amplitude on the covering space.
\end{enumerate}
\vspace{5pt}Including the factors from \ref{comp1}-\ref{comp4}, along with the base amplitude \eqref{BaseAmpResult} gives us
\begin{equation} \label{CompAmpResult}
    \widetilde{\mathcal{A}}(w_2,w_1;\phiNS) = \bigg(\frac{a}{a-b}\bigg)^{\!h}  \bigg(\frac{\sqrt{a}-\sqrt{b}}{\sqrt{a}+\sqrt{b}}\,\bigg)^{\!m} \ \frac{a-b}{4\sqrt{ab}} \ .
\end{equation}
Using \eqref{abDef} and \eqref{swDef}, this can be rewritten in terms of $w_1$ and $w_2$ as
\begin{equation}\label{CompAmpResultw}
    \widetilde{\mathcal{A}}(w_2,w_1;\phiNS) = \frac{\big(\!\cosh\!\frac{\Delta w}{4}\big)^{2h-1}}{4 \sinh\!\frac{\Delta w}{4}}\ e^{-\frac{m\Delta w}{2}} \ .
\end{equation}
From this result, the amplitude in the second term of \eqref{amp 2} can also be inferred by observing that, whilst not a superconformal primary, $G^+_{-,-\h}|\phi\rangle_{\scriptscriptstyle{N\!S}}$ is annihilated by the modes $L_n$ and $J^3_1$, since
\begin{equation}
    \begin{aligned}
        L_{n>0}\, G^+_{-,-\frac12} |\phi\rangle_{\scriptscriptstyle{N\!S}} &= \frac12(n+1) G^+_{-,n-\frac12} |\phi\rangle_{\scriptscriptstyle{N\!S}} = 0 \ ,\\
        J^3_{n>0}\, G^+_{-,-\frac12} |\phi\rangle_{\scriptscriptstyle{N\!S}} &= \frac12 G^+_{-,n-\frac12} |\phi\rangle_{\scriptscriptstyle{N\!S}} = 0 \ ,
    \end{aligned}
\end{equation}
using the algebra in \eqref{app com currents} and the superconformal primary conditions \eqref{def primary 2} on $|\phi\rangle_{\scriptscriptstyle{N\!S}}$. That these were the only conditions on $|\phi\rangle_{\scriptscriptstyle{N\!S}}$ used in steps \ref{comp1}-\ref{comp4} for the calculation of $\widetilde{\mathcal{A}}(w_2,w_1;\phiNS)$ above, allows us to obtain $\widetilde{\mathcal{A}}(w_2,w_1;G^+_{-,-\h}\phiNS)$ by simply shifting the dimension and charge in \eqref{CompAmpResultw} by $h\to h+\tfrac12$ and $m\to m+\tfrac12$ respectively\footnote{In Appendix~\ref{app.SF} we show that these conditions are sufficient for the field $G^+_{-,-\frac12}\phi$ to have the same transformation under spectral flow on the plane as the superconformal primary $\phi$.}. We now set the charge of $|\phi\rangle_{\scriptscriptstyle{N\!S}}$ to $m=0$, as explained in Section~\ref{sec:SingleSCP} and get the final result for \eqref{amp 2} as
\begin{equation} \label{AmpResult}
    A^{(1)(1)}(\Delta w) = -\frac{\big(\!\cosh\! \frac{\Delta w}{4}\big)^{2h-2}}{8\big(e^{\frac{\Delta w}{2}}-1\big)^2} \bigg[ h\, e^{\Delta w}-2(h-1)e^{\frac{\Delta w}{2}}+h \bigg] \ .
\end{equation}

\subsubsection*{Projecting onto the level-$n$ subspace}

In order to relate the result \eqref{AmpResult} to the expectation value in the lift \eqref{LiftLeft2}, the projection $\mathcal{P}$ of intermediate states onto the level-$n$ subspace should now be implemented. As discussed in Section~\ref{sec:LiftForm}, this is done by extracting the $e^{\Delta w}$-independent part of \eqref{AmpResult}. We expand in powers of $e^{-\Delta w}$ and using
\begin{equation}
\begin{aligned}
    \Big(e^{\frac{\Delta w}{2}}-1\Big)^{\!-2} &= e^{-\Delta w} \sum_{s=0}^{\infty}\, (s+1)\, e^{-s\frac{\Delta w}{2}} \ ,\\
    \bigg(\!\cosh \frac{\Delta w}{4}\bigg)^{2h-2} &= 2^{2-2h}e^{(h-1)\frac{\Delta w}{2}} \sum_{r=0}^{2h-2} {}^{2h-2}C_r \:e^{-r\frac{\Delta w}{2}} \ ,
\end{aligned}
\end{equation}
gives us the expansion of \eqref{AmpResult} as
\begin{align} \label{AmpResultExp}
    A^{(1)(1)}(\Delta w) = -\frac18 \bigg(e^{-\Delta w} \sum_{s=0}^{\infty} \,(s+1)\, e^{-s\frac{\Delta w}{2}}\bigg)& \bigg(2^{2-2h} e^{(h-1)\frac{\Delta w}{2}} \sum_{r=0}^{2h-2} {}^{2h-2}C_r \:e^{-r\frac{\Delta w}{2}}\bigg)  \nonumber\\
    &\times\bigg[\underbrace{h\,e^{\Delta w}}_{(1)} - \underbrace{2(h-1)e^{\frac{\Delta w}{2}}}_{(2)} + \underbrace{h}_{(3)}\bigg] \ .
\end{align}
For the power of $e^{-\Delta w}$ to vanish in the three terms of the square brackets of \eqref{AmpResultExp}, the summation indices in each case must satisfy
\begin{subequations} \label{eq.powers}
\begin{align}
    (1)&\quad\ s-h+r+1=0 \ \ \Longrightarrow\ \ s=h-r-1\geq0 \ \ \Longrightarrow\ \ r\leq h-1 \ ,\\
    (2)&\quad\ s-h+r+2=0 \ \ \Longrightarrow\ \ s=h-r-2\geq0 \ \ \Longrightarrow\ \ r\leq h-2 \ ,\\
    (3)&\quad\ s-h+r+3=0 \ \ \Longrightarrow\ \ s=h-r-3\geq0 \ \ \Longrightarrow\ \ r\leq h-3 \ .
\end{align}
\end{subequations}
Implementing these constraints in \eqref{AmpResultExp} and evaluating the remaining sum yields the diagonal two-copy contribution \eqref{LiftLeft2} to the lift
\begin{equation} \label{eq.E11}
    E^{(2)}_{(1)(1)}(\Phi) = \frac{\lambda^2\pi^2}{2^{2h}} \frac{\Gamma(2h)}{\Gamma(h)^2} \ .
\end{equation}
As discussed in Section~\ref{sec.states}, all of the diagonal two-copy contributions of the form $E^{(2)}_{(i)(i)}$ will be equal due to permutation invariance. However, it is not immediately clear that the off-diagonal two-copy contributions $E^{(2)}_{(j)(i\neq j)}$ will also be of the form \eqref{eq.E11}. This does in fact turn out to be the case\footnote{This same property of the various contributions to the lift was also observed in \cite{Hampton:2018ygz}.} as shown explicitly for $E^{(2)}_{(2)(1)}$ in Appendix~\ref{app.B}. Using the relation \eqref{LiftLeftN}, the full lift is then given by
\begin{equation} \label{eq.liftNew}
    E^{(2)}_{h,0}(\Phi) = (N-1)\frac{\lambda^2\pi^2}{2^{2h-1}} \frac{\Gamma(2h)}{\Gamma(h)^2} \ .
\end{equation}
The lift \eqref{eq.liftNew} interestingly does not depend on other details of the superconformal primary $\phi$, just its dimension and charge. From a technical perspective, this has happened simply due to the choice of state \eqref{FullState} -- specifically due to only exciting one copy above the Ramond ground state. Because of this, in the end we are left with a two-point function on the covering space, which does not depend on the details of the state. If more than one copy in the initial and final states was excited with the superconformal primary $\phi$ then there would be contributions to the lift where the first twist operator acts on two excited copies. These contributions would then yield a correlation function on the covering space of the schematic form $\sim\langle \phi^{\dagger}\phi^{\dagger}\phi\phi\rangle$ and hence the lift would depend on the details of the state.

One existing data point that the general formula \eqref{eq.liftNew} can be compared to was computed in \cite{Guo:2019ady}. In that paper the lifting of all states at level $1$ was computed for the case of $N=2$. One of these states is the normalised Ramond-sector state
\begin{equation} \label{stateBin}
    |\widetilde{\phi}\,\rangle_R^{(\mathcal{A})} \equiv \frac12 d_{-1}^{--(\mathcal{A})} d_0^{+-(\mathcal{A})} \Roobmk{1}\Roobmk{2} \ .
\end{equation}
The superscript $(\mathcal{A})$ denotes that these modes are anti-symmetrised over the copies in the state, \textit{i.e.} in this case of $N=2$
\begin{equation}
    d^{\alpha A(\mathcal{A})}_{n} \equiv d^{\alpha A(1)}_{n} - d^{\alpha A(2)}_{n} \ .
\end{equation}
The state \eqref{stateBin} is the lowest-weight state of a short multiplet at the free point, which is then lifted at second order in $\lambda$. In total, four short multiplets of the free theory combine into a long multiplet of the deformed theory, for which the computed lift is $E^{(2)}=\pi^2\lambda^2$. The state \eqref{stateBin} is the bottom member of this long multiplet, with the lowest $A$ charge of $-1$. While not being of the form \eqref{FullState} of states we consider here, it can be related to one by adding a global mode excitation (and normalising) to get
\begin{equation} \label{stateBin2}
    |\widetilde{\phi}\rangle_R \equiv \frac1{\sqrt{2}}\Big( |\widetilde{\phi}\,\rangle^{(\mathcal{A})} + |\widetilde{\phi}\,\rangle^{(g)} \Big) = \frac1{2\sqrt{2}}\Big( d_{-1}^{--(1)} d_0^{+-(1)} + d_{-1}^{--(2)} d_0^{+-(2)}\Big) \Roobmk{1}\Roobmk{2} \,,
\end{equation}
where these global modes are defined analogously to \eqref{globalL}. It was shown in Appendix D of \cite{Guo:2019ady} that such global modes are not lifted and so the lift of the state \eqref{stateBin2} is simply that of \eqref{stateBin} multiplied by the extra normalisation factor for both the initial and final state: that is $E^{(2)}(\widetilde{\phi}_R) =\tfrac12\pi^2\lambda^2$. The quantum numbers for the R-sector state \eqref{stateBin2} are $h_R=\tfrac32$, $m_R=-1$. Spectral flowing to the NS sector using \eqref{sf_O} with $\eta=1$, we find
\begin{equation}
    |\widetilde{\phi}\rangle_R \longrightarrow \frac1{2\sqrt{2}}\Big( d_{-\h}^{--(1)} d_{-\h}^{+-(1)}+ d_{-\h}^{--(2)} d_{-\h}^{+-(2)} \Big)\NSok{1}\NSok{2} \ ,
\end{equation}
with dimension and charge $(h,m)=(1,0)$. From \eqref{eq.liftNew} we then find
\begin{equation}
    \left. E^{(2)}_{1,0}\right|_{N=2} = \frac12\pi^2 \lambda^2 \ ,
\end{equation}
in agreement with the prediction from \cite{Guo:2019ady}.
\begin{figure}[ht]
    \centering
    \includegraphics[scale=0.9]{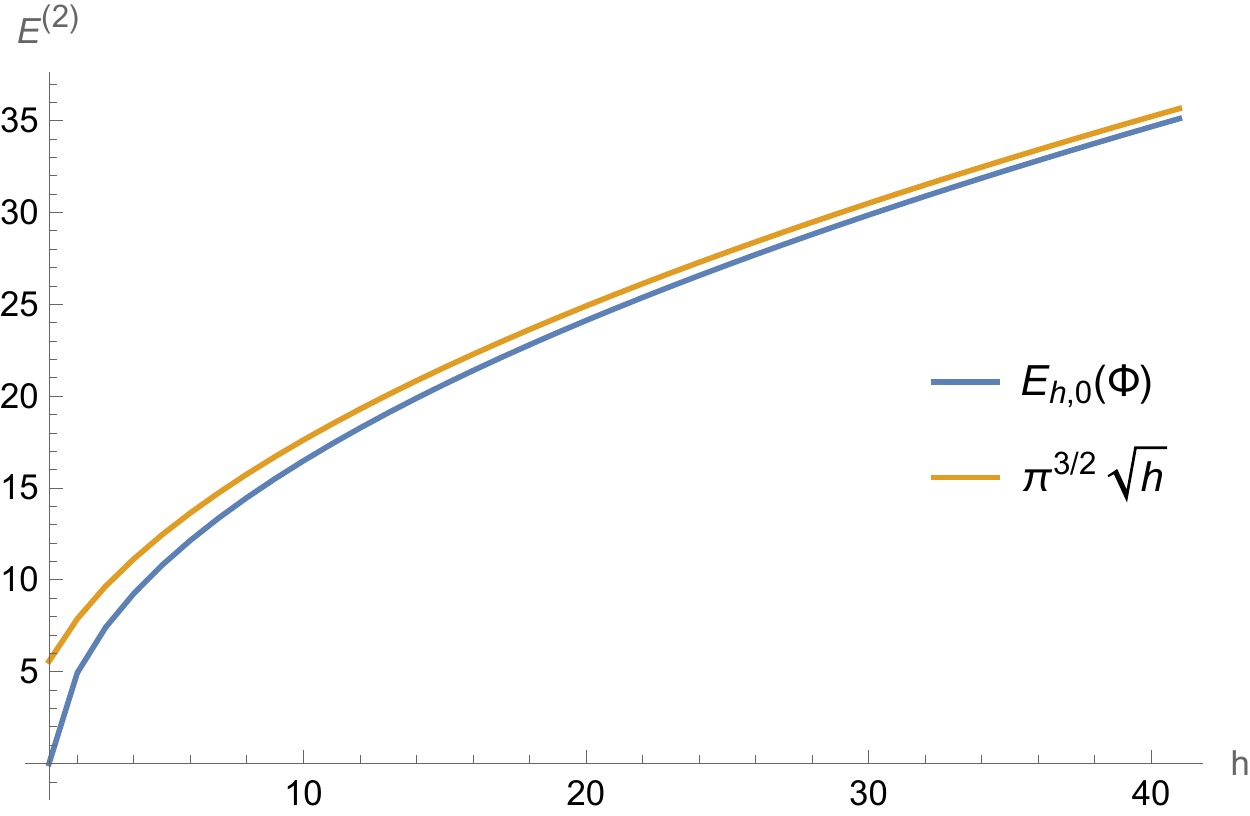}
    \caption{The second-order energy lift \eqref{eq.liftNew} for states of the form \eqref{FullState}, with $N=2$, as a function of the superconformal primary's dimension $h$. The plot shows the behavior of the lift for large $h$ and an asymptotic $\sqrt{h}$ behaviour is observed.}
    \label{fig1}
\end{figure}
It is interesting to consider the large-dimension limit of \eqref{eq.liftNew}. This appears to confront one of the motivating puzzles discussed in Section~\ref{secintro}: since the computation of Section~\ref{sec.main} is seemingly valid for arbitrarily high dimension and it appears that nothing special happens once the black hole threshold $h=\frac{c}{24}$ is crossed, where are the vast numbers of protected black hole states implied by index calculations? The asymptotic behaviour of the lift \eqref{eq.liftNew} can be easily found (see Figure~\ref{fig1} for a plot of the lift in the case of $N=2$); using Stirling's approximation for the Gamma functions, we find
\begin{equation} \label{eq.liftLim}
    E^{(2)}_{h,0}(\Phi) \approx \lambda^2 \pi^{\frac32}\sqrt{h} + O\big(h^{-\frac12}\big)\ .
\end{equation}
An interesting observation is that all previously computed examples of lifting in the literature for which the large-dimension regime is accessible also have a $\sqrt{h}$ asymptotic behaviour~\cite{Hampton:2018ygz,Guo:2021uiu}. This functional behaviour appears to be a universal property of lifting of D1-D5-P states, although with variable powers of $\pi$ in the prefactor.

\section{Discussion} \label{sec.conc}

We have studied the lifting of untwisted D1-D5-P states with one copy excited by a left-moving superconformal primary. The anomalous dimensions at second order in the deformation parameter were found in a highly compact form which, interestingly does not depend on the details of the superconformal primary used; only its dimension. The functional form of the lift \eqref{eq.liftNew} can be compactly written as $2^{-2h}$ multiplied by the reciprocal of the Euler beta function $B(h,h)$, where $h$ is the NS-sector superconformal primary's left-moving dimension. It was shown that all single-copy superconformal primaries in the untwisted sector, that are not chiral primaries, have $j=m=0$. Otherwise the lift would also depend on the value of $j$. This lifting calculation was performed using the so-called hybrid method, combining aspects of the direct computation of twist operator correlation functions and of the Gava-Narain method. This choice was made since, for this particular case, the covering space of the 4-point functions involved are only of genus $0$. From the lift \ref{eq.liftNew}, the large-dimension limit can be seen to give a growth scaling as $\sqrt{h}$. This seemingly universal asymptotic behaviour was found also in all other studies of lifting in the D1-D5 CFT for which the family of states admits a large-dimension limit~\cite{Hampton:2018ygz,Guo:2021uiu}.

This adds another family of states of the D1-D5 CFT for which the lifting pattern is known. Previous studies tend to consider either very specific restricted families of states, chosen for their relatively simple lifting computations~\cite{Gaberdiel:2015uca,Hampton:2018ygz} which allows for greater generality in twist structure and in the value of $N$, or broad sweeps of low-level states \cite{Guo:2019ady,Guo:2020gxm,Benjamin:2021zkn}, typically for small $N$ or $h$. One overarching aim of this program of forays into the lifting landscape is to gain intuition on what types of states that are BPS at the free point become lifted at a general point in moduli space, as well as the manner of their lifting. It is hoped that this intuition can be useful in the context of black hole microstates, namely: given that in the above examples, generic families of states are completely lifted irrespective of which side of the black hole threshold $h=\frac{c}{24}$ they lie, where are the states dual to supersymmetric black hole microstates? From entropy calculations it is well known that there should exist a vast number of such BPS states and identifying them exactly in the CFT spectrum should open up new tools with which to study microstates.

One immediate extension to the class of states considered in this paper is to allow for the vacuum copies to be in arbitrary Ramond-Ramond ground states. In each of the left and right sectors there are $4$ such ground states per twist sector, as given in \eqref{qgroundstates}, allowing for a potential extension to a much larger class of states. Allowing for more than one component string to be excited with a superconformal primary $\phi$ would require knowledge of four-point functions of the form $\langle \phi\phi\phi\phi\rangle$, which in general are not known. Another extension would be to instead excite a multi-wound string with winding $k$. This introduces two main complications; firstly, the lifting would require a computation involving four twist operators (two of order $k$ from the initial and final states and two of order $2$ from the deformation operators) for which the noninvertability of the covering map would in general cause problems. Secondly, the action of the first deformation operator on the initial state in the lift \eqref{lift average} can now not only twist together one unexcited copy with the excited $k$-twisted component string to produce a $(k+1)$-wound component string, but it can also split up the $k$-twisted component string. This extra type of contribution to the lift yields a genus $1$ covering space for the associated correlator, making our methods impractical for this computation. However, for large enough $N\gg k$ these new contributions are suppressed in powers of $N^{-1}$ relative to the existing terms, making this a potential avenue for exploration. We note that since the effective central charge depends on the twist sector, the arguments of Section~\ref{sec:SingleSCP} no longer hold. For higher twist sectors, there will be relevant superconformal primaries in $SU(2)$ multiplets with higher values of $j$. Thus, these more general cases will involve a larger family of states for which the lifting can be computed.

We have noted that in many instances of the lifting computation, the lift for large dimensions $h\gg1$ behaves as $\sim \sqrt{h}$ for fixed charge $m$. Let us note a physical implication of this fact. While we think of $h$ as a dimensionless number in the CFT, it actually corresponds to an energy scale
\begin{equation}
    E= \frac{h}{R_y} \ ,
\end{equation}
where $R_y$ is the radius of the $S^1$ on which the D1-D5 CFT is compactified (with the usual dimensionless energy in the CFT found by setting $R_y=1$). Now note that for $h\gg1$, the excited state has energy $E\gg 1/R_y$ and we can imagine an excitation in the CFT that is localised in a region much smaller than the infrared `box size' $2\pi R_y$. The CFT has no other energy scales -- the perturbation parameter $\lambda$ is also dimensionless -- thus if we could really ignore the `box size', then on dimensional grounds we would expect the lift to be $\Delta E \sim E$. In dimensionless units this gives a lift scaling as $\sim h$ rather than $\sim \sqrt{h}$. Since we actually find the lift to be $\sim \sqrt{h}$, we observe that \emph{the lifting process does in fact involve the box size}; the lift energy is proportional to the geometric mean of the energy $\sim h/R_y$ and the infrared scale set by the box size $1/R_y$.

\section*{Acknowledgements}

We would like to thank \'{E}tienne Ligout and Ida Zadeh for discussions. This work is supported in part by DOE grant DE-SC0011726. BG is supported by ERC Grant 787320 - QBH Structure.

\appendix

\section{The $\mathcal N=4$ superconformal algebra} \label{app_cft}

We follow the notation in the appendix A of \cite{Hampton:2018ygz}. The indices $\alpha=(+,-)$ and $\bar \alpha=(+,-)$ correspond to the subgroups $SU(2)_L$ and $SU(2)_R$ arising from rotations on $S^3$. The indices $A=(+,-)$ and $\dot A=(+,-)$ correspond to the subgroups $SU(2)_1$ and $SU(2)_2$ arising from rotations in $T^4$. We use the convention
\be
    \epsilon_{+-}=1 \quad,\quad\epsilon^{+-}=-1 \ .
\ee

\subsection{Commutation relations} \label{app.comms}

The commutation relations for the small $\mathcal N=4$ superconformal algebra are
\begin{subequations} \label{app com currents}
\begin{align}
\big[L_m,L_n\big] &= \frac{c}{12}m(m^2-1)\delta_{m+n,0}+ (m-n)L_{m+n} \ , \label{LLcomm}\\
\big[J^a_{m},J^b_{n}\big] &= \frac{c}{12}m\,\delta^{ab}\delta_{m+n,0} +  i\epsilon^{ab}_{\,\,\,\,c}\,J^c_{m+n} \ ,\label{JJcomm}\\
\big\{ G^{\alpha}_{\dot{A},r} , G^{\beta}_{\dot{B},s} \big\} &=  \epsilon_{\dot{A}\dot{B}}\bigg[\epsilon^{\alpha\beta}\frac{c}{6}\Big(r^2-\frac14\Big)\delta_{r+s,0} + \big(\sigma^{aT}\big)^{\alpha}_{\gamma}\:\epsilon^{\gamma\beta}(r-s)J^a_{r+s} + \epsilon^{\alpha\beta}L_{r+s} \bigg] \ ,\label{GGcomm}\\
\big[J^a_{m},G^{\alpha}_{\dot{A},r}\big] &= \h\big(\sigma^{aT}\big)^{\alpha}_{\beta}\, G^{\beta}_{\dot{A},m+r} \ ,\label{JGcomm}\\
\big[L_{m},G^{\alpha}_{\dot{A},r}\big] &= \Big(\frac{m}{2}  -r\Big)G^{\alpha}_{\dot{A},m+r} \ ,\label{LGcomm}\\
\big[L_{m},J^a_n\big] &= -nJ^a_{m+n} \ , \label{LJcomm}
\end{align}
\end{subequations}
with the right-moving modes satisfying an analogous set of relations. We will not have need for the full contracted large $\mathcal{N}=4$ superconformal algebra of the D1-D5 CFT, but this can nonetheless be found in Appendix A of \cite{Guo:2019ady} or Appendix A.4 of \cite{Hampton:2018ygz} with the same conventions. We do, however, give our conventions for the realisation of part of this algebra in terms of the free fermions $\psi^{\alpha A}$ with modes $d^{\alpha A}_r$, and the free bosons $\partial X_{\!A\dot{A}}$ with modes $\alpha_{A\dot{A},n}$. The mode expansions are given by
\begin{subequations} \label{BosFerModes}
    \begin{align}
        \partial X_{A\dot{A}}(z) = -i \sum_n z^{-n-1}\,\alpha_{A\dot{A},n} \ ,\\
        \psi^{\alpha A}(z) = \sum_r z^{-r-1/2}\, d^{\alpha A}_{r} \ ,
    \end{align}
\end{subequations}
and the brackets are
\begin{subequations}\label{FundComms}
    \begin{align}
        \big[\alpha_{A\dot{A},n},\alpha_{B\dot{B},m}\big] &= - n\frac{c}{6}\epsilon_{AB}\,\epsilon_{\dot{A}\dot{B}}\, \delta_{n+m,0} \ ,\label{eq.alalcomm}\\
        \big\{d^{\alpha A}_{r},d^{\beta B}_{s}\big\} &= - \frac{c}{6}\epsilon^{\alpha\beta}\epsilon^{AB} \delta_{r+s,0} \ ,\label{eq.ddcomm}
    \end{align}
\end{subequations}
and likewise for the right-moving fields.

\subsection{Relations involving the twist operator $\sigma_2^{\alpha\bar{\alpha}}$}

In the main lifting calculation of Section~\ref{sec.main}, we will have need of the following relations involving the twist operator $\sigma^{\alpha\bar{\alpha}}$ and the supersymmetry modes
\begin{equation} \label{A3}
\begin{aligned}
    G_{\dot{A},-\frac{1}{2}}^{+(0)}\sigma^{ + \bar\alpha}=0 \quad,\quad G_{\dot{A},-\frac{1}{2}}^{-(0)}\sigma^{ - \bar\alpha}=0 \ ,\\
    \bar G_{\dot{A},-\frac{1}{2}}^{+(0)}\sigma^{\alpha +}=0 \quad,\quad \bar G_{\dot{A},-\frac{1}{2}}^{-(0)}\sigma^{ \alpha -}=0 \ .
\end{aligned}
\end{equation}
On top of these, we also have the relations
\be\label{A4}
G_{\dot{A},-\frac{1}{2}}^{-(0)}\sigma^{ + \bar\alpha}=-G_{\dot{A},-\frac{1}{2}}^{+(0)}\sigma^{ - \bar\alpha} \quad,\quad
\bar G_{\dot{A},-\frac{1}{2}}^{-(0)}\sigma^{ \alpha +}=-\bar G_{\dot{A},-\frac{1}{2}}^{+(0)}\sigma^{ \alpha -} \ .
\ee
These can be proved by the following chain of logic
\be
    G_{\dot{A},-\frac{1}{2}}^{-(0)}\sigma^{ + \bar\alpha}=\Big[J^{-(0)}_{0},G_{\dot{A},-\frac{1}{2}}^{+(0)}\Big] \sigma^{ + \bar\alpha}
    = J^{-(0)}_{0}G_{\dot{A},-\frac{1}{2}}^{+(0)}\sigma^{ + \bar\alpha}-G_{\dot{A},-\frac{1}{2}}^{+(0)}J^{-(0)}_{0}\sigma^{ + \bar\alpha}
    = -G_{\dot{A},-\frac{1}{2}}^{+(0)}\sigma^{ - \bar\alpha} \ ,
\ee
where in the last step we use \eqref{A4} and $J^{-(0)}_{0}\sigma^{ + \bar\alpha}=\sigma^{ - \bar\alpha}$. In all of the above, the superscript $(0)$ refers to the fact that these are the supersymmetry modes of the undeformed theory's generators.

\subsection{Rules for Hermitian conjugation} \label{app.Herm}

Suppose we consider an amplitude on the cylinder, in the NS sector, with the form
\be
    A={}_{\scriptscriptstyle{N\!S}}\langle 0|\, {\mathcal O}^\dagger\big( \tau=T, \sigma=0 \big) {\mathcal O}\big( \tau=-T, \sigma=0 \big) |0\rangle_{\scriptscriptstyle{N\!S}} \ .
\label{appone}
\ee
Then we should have $A\ge 0$. This requirement helps determine the way Hermitian conjugates are defined in our CFT. Note that contractions between $su(2)$ indices are done using antisymmetric tensors like $\epsilon_{\alpha\beta}$, and this fact gives rise to certain negative signs in the definitions of Hermitian conjugates. For the supercharges, we use the following rules
\begin{equation} \label{Gconj}
\begin{aligned}
    \Big( G^{+}_{+}(\tau,\sigma)\Big)^{\dagger} &=-G^{-}_{-}(-\tau,\sigma)\quad,\quad \Big( G^{+}_{-}(\tau,\sigma)\Big)^{\dagger}=G^{-}_{+}(-\tau,\sigma) \ ,\\
    \Big( \bar G^{+}_{+}(\tau,\sigma)\Big)^{\dagger} &=-\bar G^{-}_{-}(-\tau,\sigma)\quad, \quad \Big( \bar G^{+}_{-}(\tau,\sigma)\Big)^{\dagger}=\bar G^{-}_{+}(-\tau,\sigma) \ ,
\end{aligned}
\end{equation}
while for the degree-2 twist operators, our conventions are
\be \label{TwistConj}
    \big(\sigma^{--}(\tau,\sigma)\big)^{\dagger}=-\sigma^{++}(-\tau,\sigma)\quad\ ,\qquad \big(\sigma^{-+}(\tau,\sigma)\big)^{\dagger}=\sigma^{+-}(-\tau,\sigma) \ .
\ee
With these choices, we find for the Gava-Narain type operators
\be \label{conju}
    \bar G^{+(P)\dagger}_{+,0}=-\bar G^{-(P)}_{-,0}\quad\ ,\qquad \bar G^{+(P)\dagger}_{-,0}=\bar G^{-(P)}_{+,0} \ .
\ee
We also use a set of `notational conventions' regarding the twist operators that drastically help in writing out the computations. A twist operator acts on both the left and right sectors, since it is a geometric deformation of the 2-dimensional spacetime on which the CFT lives. Thus strictly speaking, we cannot separate an operator like $\sigma^{\alpha \bar\alpha}$ into a $\sigma^\alpha$ for the left-mover and a $\bar\sigma^{\bar\alpha}$ for the right-mover. However, we would still like such a separation for ease of writing out expressions and to then be able to combine left- and right-moving terms consistently at the end. Attempting such a consistent separation encounters a difficulty with the signs to be used in Hermitian conjugation. This can be seen by considering the expression \eqref{appone} with ${\mathcal O}=\sigma^+$. Suppose we choose a Hermitian conjugation convention of $(\sigma^+)^\dagger=\sigma^-$. We would then find that \mbox{${}_{\scriptscriptstyle{N\!S}}\langle 0 |\, \sigma^-(T,0)\, \sigma^+(-T,0)\,|0\rangle_{\scriptscriptstyle{N\!S}}>0$}. However, since the OPE has the form
\be\label{apptwo}
    \sigma^\alpha (z)\, \sigma^\beta(0)\sim \frac{\epsilon^{\alpha\beta}}{z} \ ,
\ee
we would expect \mbox{${}_{\scriptscriptstyle{N\!S}}\langle 0|\, \sigma^+(T,0)\, \sigma^-(-T,0)\,|0\rangle_{\scriptscriptstyle{N\!S}}<0$}. This contradicts the fact that $(\sigma^-)^\dagger=\sigma^+$. It turns out that there is no consistent choice of Hermitian conjugation for a `left' part of the twist operator. For our purposes, however, we never require the use of Hermitian conjugations of the split twist operators and so we can neglect this issue here. This leaves us two options for conventions that yield consistent recombined formulae. We can either use the slightly unnatural-looking conventions
\begin{align}
    \langle \sigma^+ \sigma^- \rangle = 1 \quad,\quad \langle \bar\sigma^+ \bar\sigma^- \rangle = -1 \ ,
\end{align}
or the left/right symmetric conventions
\begin{align} \label{eq.OurConv}
    \langle \sigma^+ \sigma^- \rangle = 1 \quad,\quad \langle \bar\sigma^+ \bar\sigma^- \rangle = 1 \ .
\end{align}
We choose the latter option, however, in doing so it is also necessary to add the prescription that when the left and right parts are brought back together, an extra minus sign should be included per usage of one of the above inner products.

\section{Equality of $E^{(2)}_{(1)(1)}$ and $E^{(2)}_{(2)(1)}$} \label{app.B}

In Section~\ref{LiftComp}, in order to get the full lift $E^{(2)}_{h,0}(\Phi)$ in \eqref{eq.liftNew} from the particular diagonal two-copy contribution $E^{(2)}_{(1)(1)}$ we used that fact that an off-diagonal term contributes an equal amount. In order to show explicitly that this is the case, we consider the contribution
\begin{equation} \label{eq.offdiag}
    E^{(2)}_{(2)(1)} \equiv -2\lambda^2\pi^2\, \phiRb{2} \Romb{1} ( G^{+}_{-,-\h}\sigma^{-})\mathcal{P}( G^{-}_{+,-\h}\sigma^{+}) \phiRk{1}\Romk{2} \ .
\end{equation}
This differs from the contribution $E^{(2)}_{(1)(1)}$ computed in Section~\ref{LiftComp} by the choice of excited copy in the final state. These types of terms are what we dub `off-diagonal'. Starting from result \eqref{AmpResult} of the left-moving amplitude on the cylinder defined in \eqref{amp}, we can generate the equivalent amplitude for a final state with copy $2$ excited rather than copy $1$. That is the amplitude
\begin{equation} \label{eq.Amp21}
    A^{(2)(1)}(w_2,w_1) \equiv \phiRb{2} \Romb{1} (G^{+}_{-,-\h}\sigma^{-})(w_2)\, (G^{-}_{+,-\h}\sigma^{+})(w_1) \phiRk{1}\Romk{2} \ .
\end{equation}
This is done, as in \cite{Hampton:2018ygz}, by taking the insertion of the second deformation operator around the cylinder using $\sigma_2\to\sigma_2 + 2\pi$. This has the effect of exchanging copy $1$ and copy $2$ in the action of the second deformation operator on the doubly-wound component string produced by the action of the first deformation operator on the initial state. We therefore have to take $\Delta w\to \Delta w + 2\pi i$ in \eqref{AmpResult}: \textit{i.e.} we get
\begin{equation}
    A^{(1)(1)}(w_2,w_1)\rightarrow A^{(1)(1)}(w_2+2\pi i,w_1) = A^{(2)(1)}(w_2,w_1) \ .
\end{equation}
We note that due to the presence of the projection operator in the lift, projecting the relevant amplitude onto its $e^{w_2-w_1}$-independent term, the lift is independent of the deformation operator insertion points. Taking one deformation operator around the cylinder to generate an off-diagonal term should then not give a different contribution to the lift. Thus we expect that $E^{(2)}_{(1)(1)}$ and $E^{(2)}_{(2)(1)}$ are equal, a result that we explicitly check below. Using the result
\begin{equation}
        \bigg(\!\cosh \frac{\Delta w}{4}\bigg)^{2h-2} \to (-1)^{h-1} \bigg(\!\sinh \frac{\Delta w}{4}\bigg)^{2h-2} \ ,
\end{equation}
in \eqref{AmpResult} we get the new amplitude
\begin{equation} \label{eq.Amp21Result}
    A^{(2)(1)}(\Delta w) = (-1)^{h}\:\frac{\big(\!\sinh\! \frac{\Delta w}{4}\big)^{2h-2}}{8\big(e^{\frac{\Delta w}{2}}+1\big)^2} \bigg[h\,e^{\Delta w}+2(h-1)e^{\frac{\Delta w}{2}}+h\bigg] \ .
\end{equation}
In order to relate this to the lift, we need to project onto the level-$n$ subspace by extracting the $e^{\Delta w}$-independent part of \eqref{eq.Amp21Result}. This has the effect of implementing the projection operator $\mathcal{P}$ in \eqref{eq.offdiag}. Using the expansions in powers of $e^{-\Delta w}$
\begin{equation}
\begin{aligned}
    \Big(e^{\frac{\Delta w}{2}}+1\Big)^{\!-2} &= e^{-\Delta w} \sum_{s=0}^{\infty}\, (s+1)(-1)^{s}\, e^{-s\frac{\Delta w}{2}} \ ,\\
    \bigg(\!\sinh \frac{\Delta w}{4}\bigg)^{2h-2} &= 2^{2-2h}e^{(h-1)\frac{\Delta w}{2}} \sum_{r=0}^{2h-2} {}^{2h-2}C_r \:(-1)^{r}\, e^{-r\frac{\Delta w}{2}} \ ,
\end{aligned}
\end{equation}
gives us the expanded amplitude
\begin{align} \label{Amp21ResultExp}
    A^{(2)(1)} = \frac18 (-1)^{h} \bigg( e^{-\Delta w} \sum_{s=0}^{\infty} \,(s+1)(-1)^{s}\, &e^{-s\frac{\Delta w}{2}}\bigg) \bigg(2^{2-2h} e^{(h-1)\frac{\Delta w}{2}} \sum_{r=0}^{2h-2} {}^{2h-2}C_r \:(-1)^{r}\,e^{-r\frac{\Delta w}{2}}\bigg)\nonumber\\
    &\times\bigg[\underbrace{h\,e^{\Delta w}}_{(1)} + \underbrace{2(h-1)e^{\frac{\Delta w}{2}}}_{(2)} + \underbrace{h}_{(3)}\bigg] \ .
\end{align}
For the power of $e^{-\Delta w}$ to vanish in the three terms of the square brackets of \eqref{Amp21ResultExp}, the summation indices in each case must satisfy the same constraints as in \eqref{eq.powers}. Resumming, we get a contribution to the lift of
\begin{equation}
    E^{(2)}_{(2)(1)}(\Phi) = \frac{\lambda^2\pi^2}{2^{2h}} \frac{\Gamma(2h)}{\Gamma(h)^2} \ .
\end{equation}
which is exactly the result found for $E^{(2)}_{(1)(1)}$ in \eqref{eq.E11}.

\section{Transformation rules under spectral flow} \label{app.SF}

In this appendix we derive the conditions for an operator $\mathcal{O}(w_0)$, with $J^3$ charge $m$, to transform under a spectral flow, at $\tilde{w}$ and by $\eta$ units, as
\begin{equation} \label{sf_O0}
    \mathcal{O}(w_0)\to \big(e^{w_0}-e^{\tilde{w}}\big)^{-\eta m}\mathcal{O}(w_0)\ ,
\end{equation}
as discussed in Section~\ref{sf_O}. Given a particular representation of the $\mathcal{N}=4$ superconformal algebra \eqref{app com currents}, specified by the value of the central charge and the dimension and $J^3$ charge of the ground state, the currents will be written in terms of a set of fundamental fields $\{\mathcal{O}^{(i)}\}$. In general this will be a mixture of bosons and fermions. In this paper we have in mind the $c=6$ representation with ground-state dimension and charge $h=m=0$, in which case we have four free bosons $\partial X_{\!A\dot{A}}$ and four free fermions $\psi^{\alpha A}$ with periodicities relevant for the NS algebra. Thus in this representation $i=1,\dots,8$. Spectral flow then acts on these fundamental fields by changing the periodicity of the fermions in the manner of \eqref{sf_O0}. Since the fundamental bosons have $m=0$, we can then write that collectively for the fundamental fields
\begin{equation} \label{eq.sf_2}
    \mathcal{O}^{(i)}(w_0) \rightarrow \big(e^{w_0}-e^{\tilde{w}}\big)^{-\eta m_i} \mathcal{O}^{(i)}(w_0) \ .
\end{equation}

\subsection{Operators on the cylinder} \label{sec.SFcyl}

An arbitrary field in the spectrum $\mathcal{O}_{\mathrm{cyl}}(w_0)$ on the cylinder will be made out of a number of modes $\mathcal{O}^{(i)}_{-n_i}$ of the fundamental fields $\mathcal{O}^{(i)}_{\mathrm{cyl}}(w_0)$, schematically written in the form
\begin{equation} \label{eq.OformCyl2}
    \mathcal{O}_{\mathrm{cyl}}(w_0) = \sum_b C_{i_1,\dots,i_p}^{\,b}\, \mathcal{O}^{(i_1)}_{-n_1}\cdots\mathcal{O}^{(i_p)}_{-n_p} \ , 
\end{equation}
for some set of coefficients $C_{i_1,\dots,i_p}^{\,b}(w_0)$. The mode expansions of the fundamental fields can be written as
\begin{equation} \label{eq.Oimodes}
    \mathcal{O}^{(i)}_{\mathrm{cyl}}(w_0) = \sum_{n} (-i)^{N_i} \,w_0^{n-h_i}\, \mathcal{O}^{(i)}_{-n} \ ,
\end{equation}
with $N_i$ being $0$ or $1$ if $\mathcal{O}^{(i)}$ is fermionic or bosonic respectively. In \eqref{eq.Oimodes} $n\in\mathbb{Z}$ or $n\in\mathbb{Z}+\tfrac12$ depending on the choice\footnote{In the NS sector of the theory fermions are anti-periodic on the cylinder and periodic on the plane, with the reverse being true in the R sector. This leads to fermions having half-integer modes in the NS sector and integer modes in the R sector.} of $\mathcal{O}^{(i)}$. We will therefore look at the effect of spectral flow on a composite field by studying the effect on one term in \eqref{eq.OformCyl2}, or more precisely, on the combination of modes
\begin{equation} \label{eq.OneTerm}
    \mathcal{O}^{\{i\}}_{\{n\}} \equiv \mathcal{O}^{(i_1)}_{-n_1}\cdots\,\mathcal{O}^{(i_p)}_{-n_p} \ .
\end{equation}
A given mode in \ref{eq.OneTerm} can be written as a contour integral of its field around the insertion point $w_0$
\begin{equation} 
    \mathcal{O}^{(i)}_{-n} = \oint_{C_{w_0}} \frac{dw}{2\pi i}\, (w-w_0)^{h_i-n-1} \,i^{N_i}\, \mathcal{O}^{(i)}_{\mathrm{cyl}}(w) \ ,
\end{equation}
which transforms under spectral flow (around $w=-\infty$) by $\eta$ units as
\begin{equation}
    \mathcal{O}^{(i)}_{-n} \rightarrow \mathcal{O}^{\eta(i)}_{-n} \equiv \oint_{C_{w_0}} \frac{dw}{2\pi i}\, (w-w_0)^{h_i-n-1} \,i^{N_i}\,e^{-\eta m_i w}\, \mathcal{O}^{(i)}_{\mathrm{cyl}}(w) \ ,
\end{equation}
where \eqref{eq.sf_2} has been used with $\tilde{w}=-\infty$. This spectrally flowed mode can then be written as
\begin{align}
    \mathcal{O}^{\eta(i)}_{-n} &= e^{-\eta m_iw_0} \oint_{C_{w_0}} \frac{dw}{2\pi i}\, (w-w_0)^{h_i-n-1} \,i^{N_i}\,e^{-\eta m_i (w-w_0)}\, \mathcal{O}^{(i)}_{\mathrm{cyl}}(w) \nonumber\\
    &= e^{-\eta m_iw_0} \sum_{k\geq0} \frac{(-\eta m_i)^k}{k!} \oint_{C_{w_0}} \frac{dw}{2\pi i}\, (w-w_0)^{h_i+k-n-1}\,i^{N_i}\, \mathcal{O}^{(i)}_{\mathrm{cyl}}(w) \nonumber\\
    &= e^{-\eta m_iw_0} \sum_{k\geq0} \frac{(-\eta m_i)^k}{k!} \mathcal{O}^{(i)}_{-n+k} \ .
\end{align}
Considering now the spectral flow of the product of the $p$ modes given in \eqref{eq.OneTerm}, each mode transforms in the same manner leading to
\begin{align} \label{eq.OneTermSF}
    \mathcal{O}^{\{i\}}_{\{n\}} \rightarrow \mathcal{O}^{\eta\{i\}}_{\{n\}} &= e^{-\eta m w_0} \sum_{k_1,\dots,k_p\geq0}\, \frac{(-\eta)^{K}}{k_1!\cdots k_p!}\, m_{i_1}^{k_1}\cdots m_{i_p}^{k_p}\, \mathcal{O}^{(i_1)}_{-n_1+k_1} \cdots\,\mathcal{O}^{(i_p)}_{-n_p+k_p} \ ,
\end{align}
where $m=\sum_{j=1}^{p} m_{i_j}$ and $K= \sum_{j=1}^{p} k_j$. In order to write $\mathcal{O}^{\eta\{i\}}_{\{n\}}$ in terms of an operator acting on $\mathcal{O}^{\{i\}}_{\{n\}}$, we will have use for the action of $J^3_1$ on a mode of a fundamental field
\begin{equation} \label{eq.J31O}
    J^3_1\, \mathcal{O}^{(i)}_{-n} |0\rangle_{\scriptscriptstyle{N\!S}} = m_i\, \mathcal{O}^{(i)}_{-n+1} |0\rangle_{\scriptscriptstyle{N\!S}} \ ,
\end{equation}
which can then be easily generalised to
\begin{equation} \label{eq.J31kO}
    \big(J^3_1\big)^k\, \mathcal{O}^{(i)}_{-n} |0\rangle_{\scriptscriptstyle{N\!S}} = (m_i)^k\, \mathcal{O}^{(i)}_{-n+k} |0\rangle_{\scriptscriptstyle{N\!S}} \ .
\end{equation}
One further generalisation of \eqref{eq.J31kO} is to the case
\begin{align} \label{eq.J31kOp}
    (J^3_1)^k\, \mathcal{O}^{\{i\}}_{\{n\}}|0\rangle_{\scriptscriptstyle{N\!S}} = \big(J^3_1\big)^k\, \mathcal{O}^{(i_1)}_{-n_1}\cdots\,\mathcal{O}^{(i_p)}_{-n_p} |0\rangle_{\scriptscriptstyle{N\!S}} \ .
\end{align}
For $p=1$ this is exactly \eqref{eq.J31kO}, whereas for $p=2$ we find
\begin{align} \label{eq.J31kO2}
    \big(J^3_1\big)^k\, \mathcal{O}^{(i_1)}_{-n_1} \mathcal{O}^{(i_2)}_{-n_2} |0\rangle_{\scriptscriptstyle{N\!S}} &= \big(J^3_1\big)^{k-1}\,\Big[J^3_1, \mathcal{O}^{(i_1)}_{-n_1} \mathcal{O}^{(i_2)}_{-n_2}\Big] |0\rangle_{\scriptscriptstyle{N\!S}} \nonumber\\
    &= \big(J^3_1\big)^{k-1}\,\Big( m_{i_1} \mathcal{O}^{(i_1)}_{-n_1+1} \mathcal{O}^{(i_2)}_{-n_2} + m_{i_2} \mathcal{O}^{(i_1)}_{-n_1} \mathcal{O}^{(i_2)}_{-n_2+1}\Big) |0\rangle_{\scriptscriptstyle{N\!S}} \nonumber\\
    &= \sum_{k_1=0}^{k} {}^kC_{k_1}\, m_{i_1}^{k_1} m_{i_2}^{k-k_1}\, \mathcal{O}^{(i_1)}_{-n_1+k_1} \mathcal{O}^{(i_2)}_{-n_2+k-k_1} |0\rangle_{\scriptscriptstyle{N\!S}} \ .
\end{align}
By analogy with the multinomial theorem one finds the general $p$ case to be
\begin{equation}
    (J^3_1)^k\, \mathcal{O}^{\{i\}}_{\{n\}}|0\rangle_{\scriptscriptstyle{N\!S}} =\quad \sum_{\mathclap{k_1,\dots,k_p\geq0}}{\vphantom{\sum}}'\ \ \, \prod_{j=1}^{p} \frac{k!}{k_j!}\, m_{i_j}^{k_j}\, \mathcal{O}^{(i_j)}_{-n_j+k_j} |0\rangle_{\scriptscriptstyle{N\!S}} \ ,
\end{equation}
where the primed sum is defined as the sum over non-negative $k_1,\dots,k_p$ subject to the constraint $k_1+\dots+k_p = k$. We then immediately find that
\begin{align} \label{eq.eJ31Op}
    e^{-\eta J^3_1}\, \mathcal{O}^{\{i\}}_{\{n\}}|0\rangle_{\scriptscriptstyle{N\!S}} &= \sum_{k=0}^{\infty} \frac{(-\eta)^k}{k!} (J^3_1)^k\, \mathcal{O}^{\{i\}}_{\{n\}}|0\rangle_{\scriptscriptstyle{N\!S}} \nonumber\\
    &= \sum_{k=0}^{\infty} \frac{(-\eta)^k}{k!}\ \ \ \sum_{\mathclap{k_1,\dots,k_p\geq0}}{\vphantom{\sum}}'\ \ \, \prod_{j=1}^{p} \frac{k!}{k_j!}\, m_{i_j}^{k_j}\, \mathcal{O}^{(i_j)}_{-n_j+k_j} |0\rangle_{\scriptscriptstyle{N\!S}} \ .
\end{align}
By considering a $p$-dimensional lattice of points describing the possible values of the $k_j$, the primed sum is over a $(p-1)$-dimensional sub-lattice defined by the constraint equation $\sum_j k_j = k$ for some fixed $k$. The $(p-1)$-dimensional sub-lattices for different values of $k$ have vanishing overlap and so the union of these sub-lattices is equivalent to the $p$-dimensional lattice. We can thus make the replacement
\begin{equation}
    \sum_{k\geq0}\ \ \ \,\sum_{\mathclap{k_1,\dots,k_p\geq0}}{\vphantom{\sum}}'\ \ \,\longrightarrow\, \sum_{k_1,\dots,k_p\geq0} \ ,
\end{equation}
in \eqref{eq.eJ31Op}. By comparison with the first line of \eqref{eq.OneTermSF} we therefore find
\begin{equation} \label{eq.OneTermSF2}
    \mathcal{O}^{\{i\}}_{\{n\}} |0\rangle_{\scriptscriptstyle{N\!S}}\rightarrow \mathcal{O}^{\eta\{i\}}_{\{n\}}|0\rangle_{\scriptscriptstyle{N\!S}} = e^{-\eta m w_0} e^{-\eta J^3_1}\, \mathcal{O}^{\{i\}}_{\{n\}} |0\rangle_{\scriptscriptstyle{N\!S}} \ ,
\end{equation}
and so the field \eqref{eq.OformCyl2} on the cylinder transforms as
\begin{equation} \label{eq.OcylSF}
    \mathcal{O}_{\mathrm{cyl}}(w_0) \rightarrow e^{-\eta m w_0} e^{-\eta J^3_1}\, \mathcal{O}_{\mathrm{cyl}}(w_0) \ .
\end{equation}
Therefore, if the field on the cylinder is to transform in the manner of \eqref{sf_O0} for all $\eta$ (with $\tilde{w}= -\infty$) it is sufficient that it satisfies the condition
\begin{equation} \label{eq.CylConstraint}
    J^3_1\, \mathcal{O}_{\mathrm{cyl}}(w_0) = 0 \ .
\end{equation}

\subsection{Operators on the plane} \label{sec.SFpl}

In the lifting calculation of Section~\ref{LiftComp}, we also require the analogous condition to \eqref{eq.CylConstraint} for the plane. This is because once the component amplitudes in \eqref{amp 2} are mapped to the covering space (the $t$-plane) we are left with amplitudes of the form $\langle \mathcal{O} S^- S^+ \mathcal{O}\rangle$ with $S^{\pm}$ being spin fields. These spin field insertions can then be resolved by appropriate spectral flows around each of them. Knowledge of the transformation of the operators $\mathcal{O}$ under spectral flow is then necessary.

Instead of finding the most general condition for a field in the spectrum to transform with a simple phase factor, we focus on the two special cases required in the main body of this paper. Those cases are when $\mathcal{O}$ is either a superconformal primary $\phi$ or is of the form $G^+_{-,-\frac12}\phi$. We proceed in the following steps, firstly for $\mathcal{O}=\phi$:
\begin{enumerate}[start=1,
    labelindent=\parindent,
    leftmargin =1.25\parindent,
    label=(\arabic*)]
\item \label{step1} \vspace{5pt}
    The correlator on the $t$-plane $\langle \phi(\infty)S^-(t_2)S^+(t_1) \phi(-a)\rangle$ is mapped to the $t'$-plane, where the two are related by
    \begin{equation}
        t' = \frac{t-t_1}{t-t_2} \quad,\quad \frac{dt'}{dt} = \frac{t_1-t_2}{(t-t_2)^2} = \frac{(t'-1)^2}{t_1-t_2} \ .
    \end{equation}
    This gives the correlator $f_1 \langle \phi(1)S^-(\infty)S^+(0) \phi(T)\rangle$.
    
\item \label{step2}
    The $t'$-plane is mapped to the $w'$-cylinder, with
    \begin{equation}
        t' = e^{w'} \ ,
    \end{equation}
    giving the correlator $f_1f_2\langle \phi(0)S^-(\infty)S^+(-\infty) \phi(\log T)\rangle$.
    
\item \label{step3}
    Due to the insertions of the spin fields on the cylinder being at $w'=\pm\infty$, they are equivalent to the whole cylinder being in the Ramond sector. Now on the $w'$-cylinder we spectral flow by $\eta=-1$ at $w=-\infty$. This gives the correlator $f_1f_2f_3\langle \phi(0) \phi(\log T)\rangle$.
    
\item \label{step4}
    We now map back to the $t'$-plane yielding $f_1f_2f_3f_4\langle \phi(1) \phi(T)\rangle$.
    
\item \label{step5}
    And then to the original $t$-plane giving the correlator $f_1f_2f_3f_4f_5\langle \phi(\infty) \phi(-a)\rangle$.
    
\item \label{step6}
    This final correlator can then be compared with the `na\"{\i}ve' plane spectral flow transformation of the $t$-plane correlator. For a spectral flow around $t=\tilde{t}$ by $\eta$ units, this would be
    \begin{align} \label{eq.sftpl}
        \mathcal{O}(t_0) \rightarrow (t_0-\tilde{t}\,)^{-\eta m} \mathcal{O}(t_0) \ ,
    \end{align}
    by analogy with \eqref{sf_O0}. For this $t$-plane correlator, we would need a spectral flow by $\eta=-1$ around $t=t_1$ under which, using \eqref{eq.sftpl}, the operators transform as
    \begin{subequations}
        \begin{align}
        S^+(t_1) &\rightarrow \mathbb{I} \ ,\\
        S^-(t_2) &\rightarrow (t_2-t_1)^{-1/2} S^-(t_2) \ ,\\
        \phi(-a) &\rightarrow (-a-t_1)^m \phi(-a) \ ,\\
        \phi(\infty) &=\lim_{t\to\infty} t^{2h} \phi(t) \rightarrow \lim_{t\to\infty} t^{2h} (t-t_1)^{m} \phi(t) \ .
    \end{align}
    \end{subequations}
    This is then followed by a spectral flow by $\eta=+1$ around $t=t_2$ under which
    \begin{subequations}
        \begin{align}
        S^-(t_2) &\rightarrow \mathbb{I} \ ,\\
        \phi(-a) &\rightarrow (-a-t_2)^{-m} \phi(-a) \ ,\\
        \phi(t) &\rightarrow (t-t_2)^{-m} \phi(t) \ ,
    \end{align}
    \end{subequations}
    giving, in total
    \begin{align} \label{eq.NaiPhi}
        \langle \phi(\infty)S^-(t_2)S^+(t_1) \phi(-a)\rangle \longrightarrow& \lim_{t\to\infty}  (t_2-t_1)^{-\tfrac{1}{2}}\, t^{2h} \frac{(-a-t_1)^m}{(-a-t_2)^m}  \frac{(t-t_1)^{m}}{(t-t_2)^{m}} \langle\phi(t) \phi(-a)\rangle \nonumber\\
        &= (t_2-t_1)^{-\tfrac{1}{2}} \bigg(\frac{a+t_1}{a+t_2}\bigg)^{\!m} \langle\phi(\infty) \phi(-a)\rangle \ .
    \end{align}
\end{enumerate}
We proceed through steps \ref{step1} to \ref{step5} above and compare the result with \eqref{eq.NaiPhi} in step \ref{step6}. We require the result that a conformal primary of dimension $h$ transforms under the map $z\to z'$ homogeneously as
\begin{equation} \label{eq.primaryTrans}
    \mathcal{O}(z) \rightarrow \widetilde{\mathcal{O}}(z') = \bigg(\frac{dz'}{dz}\bigg)^{\!-h} \mathcal{O}(z) \ ,
\end{equation}
and that a spin field transforms like a dimension $h=\frac14$ primary. For ease of notation we define $\widetilde{\mathcal{O}}(z')\equiv\mathcal{O}(z')$. Therefore, one finds that from the map to the $t'$-plane in step~\ref{step1} we have the transformations
\begin{subequations}
    \begin{align}
    \phi(t=\infty) &= \lim_{t\to\infty} t^{2h} \frac{(t_1-t_2)^h}{(t-t_2)^{2h}}\,\phi(t'=1) = (t_1-t_2)^h \phi(t'=1) \ ,\\
    \phi(t=-a) &= \frac{(t_1-t_2)^h}{(-a-t_2)^{2h}}\,\phi(t'=T) \ ,\\
    S^-(t_2) &= \lim_{t'\to\infty} \bigg(\frac{(t'-1)^2}{t_1-t_2}\bigg)^{\!\frac{1}{4}} S^-(t') = \lim_{t'\to\infty} (t')^{\frac12} (t_2-t_1)^{-\frac14}(-1)^{\frac14} S^-(t') \nonumber\\&\hspace{4.62cm}= (t_2-t_1)^{-\frac14}(-1)^{\frac14} S^-(\infty) \ ,\\
    S^+(t_1) &= \lim_{t\to t_1} \bigg(\frac{t_1-t_2}{(t-t_2)^2}\bigg)^{\!\frac{1}{4}} S^+(t'=0) = (t_2-t_1)^{-\frac14}(-1)^{-\frac14} S^+(t'=0) \ ,
\end{align}
\end{subequations}
where $T\equiv \frac{a+t_1}{a+t_2}$. In total this step gives the factor
\begin{equation}
    f_1 = \frac{(t_2-t_1)^{2h-\frac12}}{(a+t_2)^{2h}} \ .
\end{equation}
In step \ref{step2}, the map to the $w'$-cylinder we have the transformations
\begin{subequations}
    \begin{align}
    \phi(t'=1) &= \phi(w'=0) \ ,\\
    \phi(t'=T) &= T^{-h} \phi(w'=\log T) \ ,\\
    S^-(t'=\infty) &= \lim_{t',w'\to\infty} (t')^{\frac12} (t')^{-\frac14} S^-(w') = \lim_{w'\to\infty} \big(e^{w'}\big)^{\frac14} S^-(w') = \lim_{w'\to\infty} \big(e^{w'}\big)^{-\frac14} S^-(w'=\infty) \ ,\\
    S^+(t'=0) &= \lim_{t'\to\infty} (t')^{-\frac14} S^+(w') = \lim_{w'\to-\infty} \big(e^{w'}\big)^{-\frac14} S^+(w') = \lim_{w'\to\infty} \big(e^{w'}\big)^{\frac14} S^+(w'=-\infty) \ .
\end{align}
\end{subequations}
So we have the overall factor from step~\ref{step2} as
\begin{equation}
    f_2 = T^{-h} \ .
\end{equation}
We now spectral flow on the cylinder by $\eta=-1$ around $w'=-\infty$, removing the spin fields and giving the transformations (using \eqref{sf_O0})
\begin{subequations}
    \begin{align}
    \phi(w'=0) &\rightarrow \phi(w'=0) \ ,\\
    \phi(w'=\log T) &\rightarrow e^{m\log T} \phi(w'=\log T) = T^{m} \phi(w'=\log T) \ .
\end{align}
\end{subequations}
These transformations are justified because $J^3_1 \phi = 0$ on the cylinder for a superconformal primary. Therefore we have the step~\ref{step3} factor
\begin{equation}
    f_3 = T^m \ .
\end{equation}
Mapping back to the $t'$-plane we get
\begin{subequations}
    \begin{align}
    \phi(w'=0) &= \lim_{w'\to0} (t')^{h} \phi(t') = \phi(t'=1) \ ,\\
    \phi(w'=\log T) &= T^h \phi(t'=T) \ ,
\end{align}
\end{subequations}
and so the step~\ref{step4} factor
\begin{equation}
    f_4 = T^h \ .
\end{equation}
Finally mapping to the $t$-plane gives us
\begin{subequations}
    \begin{align}
    \phi(t'=1) &= \lim_{t'\to1} \bigg(\frac{(t-t_2)^2}{t_1-t_2}\bigg)^{\!h} \phi(t) = \lim_{t\to\infty} t^{2h} (t_1-t_2)^{-h} \phi(t) = (t_1-t_2)^{-h} \phi(t=\infty) \ ,\\
    \phi(t'=T) &= \bigg(\frac{(-a-t_2)^2}{t_1-t_2}\bigg)^{\!h} \phi(t=-a) \ ,
\end{align}
\end{subequations}
giving the step~\ref{step5} factor
\begin{equation}
    f_5 = \bigg(\frac{-a-t_2}{t_1-t_2}\bigg)^{\!2h} \ .
\end{equation}
Multiplying the factors from each step, we have
\begin{equation}
    f_1f_2f_3f_4f_5 = \frac{(t_2-t_1)^{2h-\frac12}}{(a+t_2)^{2h}} T^{-h} T^m T^h \bigg(\frac{-a-t_2}{t_1-t_2}\bigg)^{\!2h} = T^m (t_2-t_1)^{-\frac12} = \bigg(\frac{a+t_1}{a+t_2}\bigg)^{\!m} (t_2-t_1)^{-\frac12} .
\end{equation}
Thus we see that this is precisely equal to the `na\"{\i}ve' $t$-plane spectral flow of the correlator from step~\ref{step6} above. This then concludes that: a superconformal primary transforms under spectral flow on the plane by the simple phase relation~\ref{eq.sftpl}.\\
\\
In our paper we also need the transformation rule for $G^+_{-,-\frac12}\phi$ which is not a superconformal primary but does satisfy $J^3_1\, G^+_{-,-\frac12}\phi =0$ on the cylinder. So we can again use its cylinder transformation under spectral flow to derive its plane transformation. It is first important to start on the $w$-cylinder and map to the $z$-plane via $w\to z(w)=e^w$ and then from the $z$-plane to the $t$-plane in order to check that the initial and final states on the $w$-cylinder yield two insertions of $G^+_{-,-\frac12}\phi$ on the covering space. Under the map $w\to z$, using \eqref{eq.primaryTrans} we have
\begin{align}
    \big(G^+_{-,-\frac12}\phi\big)(w=-\infty) = \lim_{w\to-\infty} \oint_{C_w}\frac{d\tilde{w}}{2\pi i}\,G^+_-(\tilde{w})\phi(w) &= \lim_{z\to0} z^h \oint_{C_z}\frac{d\tilde{z}}{2\pi i}\,\tilde{z}^{\frac12}G^+_-(\tilde{z})\phi(z) \nonumber\\ &= \lim_{z\to0} z^{h+\frac12} \big(G^+_{-,-\frac12}\phi\big)(z=0) \ ,\\
    \big(G^-_{+,\frac12}\phi\big)(w=\infty) = \lim_{w\to\infty} \big(e^w\big)^{2h+1}\! \oint_{C_w}\frac{d\tilde{w}}{2\pi i}\,G^+_-(\tilde{w})\phi(w) &= \lim_{z\to\infty} z^{3h+1} \oint_{C_z}\frac{d\tilde{z}}{2\pi i}\, \tilde{z}^{\frac12} G^+_-(\tilde{z})\phi(z) \nonumber\\&= \lim_{z\to\infty} z^{h+\frac12} \big(G^-_{+,\frac12}\phi\big)(z=\infty) \ .
\end{align}
Thus when mapping the correlator from the $w$-cylinder to the $z$-plane the zero and diverging factors above cancel and the $G$ modes remain the same. Then under the map $z\to t(z)$, where
\begin{equation}
    z(t) = \frac{(t+a)(t+b)}{t} \quad,\quad \frac{dz}{dt} = \frac{(t-t_1)(t-t_2)}{t^2} \ ,
\end{equation}
with $t_1=-\sqrt{ab}$ and $t_2=\sqrt{ab}$, we have firstly (the fields are taken to be on the first sheet, $z=0^{(1)}$ and $z=\infty^{(1)}$ respectively)
\begin{align}
    \big(G^+_{-,-\frac12}\phi\big)(z=0^{(1)}) &= \lim_{t\to-a} \bigg(\frac{dt}{dz}\bigg)^{\!h} \oint_{C_t}\frac{d\tilde{t}}{2\pi i}\,\bigg(\frac{\tilde{t}^{\,2}}{(\tilde{t}-t_1)(\tilde{t}-t_2)}\bigg)^{\!\frac12} G^+_-(\tilde{t})\phi(t) \nonumber\\
    &= \lim_{t\to-a} \frac{a^2}{(a+t_1)^h(a+t_2)^h} \oint_{C_t}\frac{d\tilde{t}}{2\pi i}\,\frac{t}{\sqrt{t^2-ab}}\Big[ 1 + O(\tilde{t}-t) \Big] G^+_-(\tilde{t})\phi(t) \nonumber\\
    &= \frac{-a^3}{a^{h+\frac12}(a-b)^{h+\frac12}} \big(G^+_{-,-\frac12}\phi\big)(t=-a) \ ,
\end{align}
and secondly
\begin{align}
    \big(G^-_{+,\frac12}\phi\big)(z=\infty^{(1)}) &= \lim_{t\to\infty} \big(z(t)\big)^{2h+1} \bigg(\frac{dt}{dz}\bigg)^{\!h} \oint_{C_t}\frac{d\tilde{t}}{2\pi i}\,\bigg(\frac{d\tilde{t}}{d\tilde{z}}\bigg)^{\!\frac12} G^+_-(\tilde{t})\phi(t) \nonumber\\
    &= \lim_{t\to\infty} \frac{[(t+a)(t+b)]^{2h+1}}{t[(t-t_1)(t-t_2)]^h}\oint_{C_t}\frac{d\tilde{t}}{2\pi i}\, \bigg(\frac{\tilde{t}^{\,2}}{(\tilde{t}-t_1)(\tilde{t}-t_2)}\bigg)^{\!\frac12} G^+_-(\tilde{t})\phi(t) \nonumber\\
    &= \lim_{t\to\infty} t^{2h+1}\oint_{C_t}\frac{d\tilde{t}}{2\pi i}\, \frac{t}{\sqrt{t^2-ab}}\Big[ 1 + O(\tilde{t}-t) \Big] G^+_-(\tilde{t})\phi(t) \nonumber\\
    &= \big(G^-_{+,\frac12}\phi\big)(t=\infty) \ .
\end{align}
So on the $t$-plane we still have only the one $G$ mode around each $\phi$. This is clearly a special case due to $\phi$ being a superconformal primary. Now we can repeat the steps~\ref{step1} to \ref{step6} of the argument above with $G\phi$ instead of $\phi$. From the case done in detail above, we see that all of the Jacobians picked up by the two $\phi$ cancel. The same will hold for $G\phi$ since it is also a conformal primary and will have just a shifted Jacobian \eqref{eq.primaryTrans} with $h\to h+\frac12$. The Jacobians and spectral flow transformations of the spin fields will be the same as before, leaving us just to calculate the spectral flow transformations of the $G\phi$ on the $w'$-cylinder.

Since $J^3_1\,G^+_{-,-\frac12}\phi=0$ we can again use the simple transformation on the cylinder under spectral flow, giving for step~\ref{step3}
\begin{align}
    \big(G^+_{-,-\frac12}\phi\big)(w'=0) &\rightarrow \big(G^+_{-,-\frac12}\phi\big)(w'=0) \ ,\\
    \big(G^+_{-,-\frac12}\phi\big)(w'=\log T) &\rightarrow e^{(m+\frac12)\log T} \big(G^+_{-,-\frac12}\phi\big)(w'=\log T) = T^{m+\frac12} \big(G^+_{-,-\frac12}\phi\big)(w'=\log T) \ .
\end{align}
Putting all of the factors together gives the $t$-plane correlator
\begin{align}
    \langle \big(G^-_{+,\frac12}\phi\big)(\infty) S^-(t_2) S^+(t_1)& \big(G^+_{-,-\frac12}\phi\big)(-a) \rangle = T^{m+\frac12} (t_2-t_1)^{-\frac12} \langle \big(G^-_{+,\frac12}\phi\big)(\infty)\big(G^+_{-,-\frac12}\phi\big)(-a)\rangle \nonumber\\
    &= \bigg(\frac{a+t_1}{a+t_2}\bigg)^{\!m+\frac12} (t_2-t_1)^{-\frac12} \langle \big(G^-_{+,\frac12}\phi\big)(\infty)\big(G^+_{-,-\frac12}\phi\big)(-a)\rangle \ ,
\end{align}
which is again exactly the `na\"{\i}ve' spectral flow on the plane given by \eqref{eq.NaiPhi} with $m\to m+\frac12$. This then concludes that: for a superconformal primary $\phi$, the field $G^+_{-,-\frac12}\phi$ also transforms under spectral flow on the plane by the simple phase relation \eqref{eq.sftpl}.

\bibliographystyle{JHEP}
\bibliography{UniversalLift.bib}

\providecommand{\href}[2]{#2}\begingroup\raggedright\begin{thebibliography}{10}

\bibitem{Strominger:1996sh}
A.~Strominger and C.~Vafa, \emph{{Microscopic origin of the Bekenstein-Hawking
  entropy}}, \href{https://doi.org/10.1016/0370-2693(96)00345-0}{\emph{Phys.
  Lett. B} {\bfseries 379} (1996) 99}
  [\href{https://arxiv.org/abs/hep-th/9601029}{{\ttfamily hep-th/9601029}}].

\bibitem{Maldacena:1999bp}
J.M.~Maldacena, G.W.~Moore and A.~Strominger, \emph{{Counting BPS black holes
  in toroidal Type II string theory}},
  \href{https://arxiv.org/abs/hep-th/9903163}{{\ttfamily hep-th/9903163}}.

\bibitem{Vafa:1995bm}
C.~Vafa, \emph{{Instantons on D-branes}},
  \href{https://doi.org/10.1016/0550-3213(96)00075-2}{\emph{Nucl. Phys. B}
  {\bfseries 463} (1996) 435}
  [\href{https://arxiv.org/abs/hep-th/9512078}{{\ttfamily hep-th/9512078}}].

\bibitem{Dijkgraaf:1998gf}
R.~Dijkgraaf, \emph{{Instanton strings and hyperKahler geometry}},
  \href{https://doi.org/10.1016/S0550-3213(98)00869-4}{\emph{Nucl. Phys. B}
  {\bfseries 543} (1999) 545}
  [\href{https://arxiv.org/abs/hep-th/9810210}{{\ttfamily hep-th/9810210}}].

\bibitem{Seiberg:1999xz}
N.~Seiberg and E.~Witten, \emph{{The D1 / D5 system and singular CFT}},
  \href{https://doi.org/10.1088/1126-6708/1999/04/017}{\emph{JHEP} {\bfseries
  04} (1999) 017} [\href{https://arxiv.org/abs/hep-th/9903224}{{\ttfamily
  hep-th/9903224}}].

\bibitem{Larsen:1999uk}
F.~Larsen and E.J.~Martinec, \emph{{U(1) charges and moduli in the D1 - D5
  system}}, \href{https://doi.org/10.1088/1126-6708/1999/06/019}{\emph{JHEP}
  {\bfseries 06} (1999) 019}
  [\href{https://arxiv.org/abs/hep-th/9905064}{{\ttfamily hep-th/9905064}}].

\bibitem{Arutyunov:1997gi}
G.E.~Arutyunov and S.A.~Frolov, \emph{{Four graviton scattering amplitude from
  S**N R**8 supersymmetric orbifold sigma model}},
  \href{https://doi.org/10.1016/S0550-3213(98)00326-5}{\emph{Nucl. Phys. B}
  {\bfseries 524} (1998) 159}
  [\href{https://arxiv.org/abs/hep-th/9712061}{{\ttfamily hep-th/9712061}}].

\bibitem{Arutyunov:1997gt}
G.E.~Arutyunov and S.A.~Frolov, \emph{{Virasoro amplitude from the S**N R**24
  orbifold sigma model}},
  \href{https://doi.org/10.1007/BF02557107}{\emph{Theor. Math. Phys.}
  {\bfseries 114} (1998) 43}
  [\href{https://arxiv.org/abs/hep-th/9708129}{{\ttfamily hep-th/9708129}}].

\bibitem{Jevicki:1998bm}
A.~Jevicki, M.~Mihailescu and S.~Ramgoolam, \emph{{Gravity from CFT on S**N(X):
  Symmetries and interactions}},
  \href{https://doi.org/10.1016/S0550-3213(00)00147-4}{\emph{Nucl. Phys. B}
  {\bfseries 577} (2000) 47}
  [\href{https://arxiv.org/abs/hep-th/9907144}{{\ttfamily hep-th/9907144}}].

\bibitem{David:2002wn}
J.R.~David, G.~Mandal and S.R.~Wadia, \emph{{Microscopic formulation of black
  holes in string theory}},
  \href{https://doi.org/10.1016/S0370-1573(02)00271-5}{\emph{Phys. Rept.}
  {\bfseries 369} (2002) 549}
  [\href{https://arxiv.org/abs/hep-th/0203048}{{\ttfamily hep-th/0203048}}].

\bibitem{Bena:2022ldq}
I.~Bena, E.J.~Martinec, S.D.~Mathur and N.P.~Warner, \emph{{Snowmass White
  Paper: Micro- and Macro-Structure of Black Holes}},
  \href{https://arxiv.org/abs/2203.04981}{{\ttfamily 2203.04981}}.

\bibitem{Bena:2022rna}
I.~Bena, E.J.~Martinec, S.D.~Mathur and N.P.~Warner, \emph{{Fuzzballs and
  Microstate Geometries: Black-Hole Structure in String Theory}},
  \href{https://arxiv.org/abs/2204.13113}{{\ttfamily 2204.13113}}.

\bibitem{Lunin:2001jy}
O.~Lunin and S.D.~Mathur, \emph{{AdS / CFT duality and the black hole
  information paradox}},
  \href{https://doi.org/10.1016/S0550-3213(01)00620-4}{\emph{Nucl. Phys. B}
  {\bfseries 623} (2002) 342}
  [\href{https://arxiv.org/abs/hep-th/0109154}{{\ttfamily hep-th/0109154}}].

\bibitem{Mathur:2005zp}
S.D.~Mathur, \emph{{The Fuzzball proposal for black holes: An Elementary
  review}}, \href{https://doi.org/10.1002/prop.200410203}{\emph{Fortsch. Phys.}
  {\bfseries 53} (2005) 793}
  [\href{https://arxiv.org/abs/hep-th/0502050}{{\ttfamily hep-th/0502050}}].

\bibitem{Kanitscheider:2007wq}
I.~Kanitscheider, K.~Skenderis and M.~Taylor, \emph{{Fuzzballs with internal
  excitations}},
  \href{https://doi.org/10.1088/1126-6708/2007/06/056}{\emph{JHEP} {\bfseries
  06} (2007) 056} [\href{https://arxiv.org/abs/0704.0690}{{\ttfamily
  0704.0690}}].

\bibitem{Bena:2007kg}
I.~Bena and N.P.~Warner, \emph{{Black holes, black rings and their
  microstates}}, \href{https://doi.org/10.1007/978-3-540-79523-0_1}{\emph{Lect.
  Notes Phys.} {\bfseries 755} (2008) 1}
  [\href{https://arxiv.org/abs/hep-th/0701216}{{\ttfamily hep-th/0701216}}].

\bibitem{Chowdhury:2010ct}
B.D.~Chowdhury and A.~Virmani, \emph{{Modave Lectures on Fuzzballs and Emission
  from the D1-D5 System}},  in \emph{{5th Modave Summer School in Mathematical
  Physics}}, 1, 2010 [\href{https://arxiv.org/abs/1001.1444}{{\ttfamily
  1001.1444}}].

\bibitem{Shigemori:2020yuo}
M.~Shigemori, \emph{{Superstrata}},
  \href{https://doi.org/10.1007/s10714-020-02698-8}{\emph{Gen. Rel. Grav.}
  {\bfseries 52} (2020) 51} [\href{https://arxiv.org/abs/2002.01592}{{\ttfamily
  2002.01592}}].

\bibitem{Gava:2002xb}
E.~Gava and K.S.~Narain, \emph{{Proving the PP wave / CFT(2) duality}},
  \href{https://doi.org/10.1088/1126-6708/2002/12/023}{\emph{JHEP} {\bfseries
  12} (2002) 023} [\href{https://arxiv.org/abs/hep-th/0208081}{{\ttfamily
  hep-th/0208081}}].

\bibitem{Gaberdiel:2015uca}
M.R.~Gaberdiel, C.~Peng and I.G.~Zadeh, \emph{{Higgsing the stringy higher spin
  symmetry}}, \href{https://doi.org/10.1007/JHEP10(2015)101}{\emph{JHEP}
  {\bfseries 10} (2015) 101}
  [\href{https://arxiv.org/abs/1506.02045}{{\ttfamily 1506.02045}}].

\bibitem{Hampton:2018ygz}
S.~Hampton, S.D.~Mathur and I.G.~Zadeh, \emph{{Lifting of D1-D5-P states}},
  \href{https://doi.org/10.1007/JHEP01(2019)075}{\emph{JHEP} {\bfseries 01}
  (2019) 075} [\href{https://arxiv.org/abs/1804.10097}{{\ttfamily
  1804.10097}}].

\bibitem{Guo:2019ady}
B.~Guo and S.D.~Mathur, \emph{{Lifting of level-1 states in the D1D5 CFT}},
  \href{https://doi.org/10.1007/JHEP03(2020)028}{\emph{JHEP} {\bfseries 03}
  (2020) 028} [\href{https://arxiv.org/abs/1912.05567}{{\ttfamily
  1912.05567}}].

\bibitem{Guo:2020gxm}
B.~Guo and S.D.~Mathur, \emph{{Lifting at higher levels in the D1D5 CFT}},
  \href{https://doi.org/10.1007/JHEP11(2020)145}{\emph{JHEP} {\bfseries 11}
  (2020) 145} [\href{https://arxiv.org/abs/2008.01274}{{\ttfamily
  2008.01274}}].

\bibitem{Lima:2020boh}
A.A.~Lima, G.M.~Sotkov and M.~Stanishkov, \emph{{Microstate Renormalization in
  Deformed D1-D5 SCFT}},
  \href{https://doi.org/10.1016/j.physletb.2020.135630}{\emph{Phys. Lett. B}
  {\bfseries 808} (2020) 135630}
  [\href{https://arxiv.org/abs/2005.06702}{{\ttfamily 2005.06702}}].

\bibitem{Lima:2020kek}
A.A.~Lima, G.M.~Sotkov and M.~Stanishkov, \emph{{Renormalization of twisted
  Ramond fields in D1-D5 SCFT$_{2}$}},
  \href{https://doi.org/10.1007/JHEP03(2021)202}{\emph{JHEP} {\bfseries 03}
  (2021) 202} [\href{https://arxiv.org/abs/2010.00172}{{\ttfamily
  2010.00172}}].

\bibitem{Lima:2020nnx}
A.A.~Lima, G.M.~Sotkov and M.~Stanishkov, \emph{{Correlation functions of
  composite Ramond fields in deformed D1-D5 orbifold SCFT$_2$}},
  \href{https://doi.org/10.1103/PhysRevD.102.106004}{\emph{Phys. Rev. D}
  {\bfseries 102} (2020) 106004}
  [\href{https://arxiv.org/abs/2006.16303}{{\ttfamily 2006.16303}}].

\bibitem{Lima:2020urq}
A.A.~Lima, G.M.~Sotkov and M.~Stanishkov, \emph{{Dynamics of R-neutral Ramond
  fields in the D1-D5 SCFT}},
  \href{https://arxiv.org/abs/2012.08021}{{\ttfamily 2012.08021}}.

\bibitem{Lima:2021wrz}
A.A.~Lima, G.M.~Sotkov and M.~Stanishkov, \emph{{On the Dynamics of Protected
  Ramond Ground States in the D1-D5 CFT}},
  \href{https://arxiv.org/abs/2103.04459}{{\ttfamily 2103.04459}}.

\bibitem{Lima:2021xqj}
A.A.~Lima, G.M.~Sotkov and M.~Stanishkov, \emph{{Ramond States of the D1-D5 CFT
  away from the free orbifold point}},  in \emph{{14th International Workshop
  on Lie Theory and Its Applications in Physics}}, 12, 2021
  [\href{https://arxiv.org/abs/2112.10832}{{\ttfamily 2112.10832}}].

\bibitem{Lima:2022cnq}
A.A.~Lima, G.M.~Sotkov and M.~Stanishkov, \emph{{Four-point functions with
  multi-cycle fields in symmetric orbifolds and the D1-D5 CFT}},
  \href{https://arxiv.org/abs/2202.12424}{{\ttfamily 2202.12424}}.

\bibitem{Benjamin:2021zkn}
N.~Benjamin, C.A.~Keller and I.G.~Zadeh, \emph{{Lifting 1/4-BPS states in
  $AdS_{3}\times S^{3}\times T^{4}$}},
  \href{https://doi.org/10.1007/JHEP10(2021)089}{\emph{JHEP} {\bfseries 10}
  (2021) 089} [\href{https://arxiv.org/abs/2107.00655}{{\ttfamily
  2107.00655}}].

\bibitem{Gaberdiel:2018rqv}
M.R.~Gaberdiel and R.~Gopakumar, \emph{{Tensionless string spectra on
  AdS$_{3}$}}, \href{https://doi.org/10.1007/JHEP05(2018)085}{\emph{JHEP}
  {\bfseries 05} (2018) 085}
  [\href{https://arxiv.org/abs/1803.04423}{{\ttfamily 1803.04423}}].

\bibitem{Eberhardt:2018ouy}
L.~Eberhardt, M.R.~Gaberdiel and R.~Gopakumar, \emph{{The Worldsheet Dual of
  the Symmetric Product CFT}},
  \href{https://doi.org/10.1007/JHEP04(2019)103}{\emph{JHEP} {\bfseries 04}
  (2019) 103} [\href{https://arxiv.org/abs/1812.01007}{{\ttfamily
  1812.01007}}].

\bibitem{Eberhardt:2019ywk}
L.~Eberhardt, M.R.~Gaberdiel and R.~Gopakumar, \emph{{Deriving the
  AdS$_{3}$/CFT$_{2}$ correspondence}},
  \href{https://doi.org/10.1007/JHEP02(2020)136}{\emph{JHEP} {\bfseries 02}
  (2020) 136} [\href{https://arxiv.org/abs/1911.00378}{{\ttfamily
  1911.00378}}].

\bibitem{Eberhardt:2020akk}
L.~Eberhardt, \emph{{AdS$_{3}$/CFT$_{2}$ at higher genus}},
  \href{https://doi.org/10.1007/JHEP05(2020)150}{\emph{JHEP} {\bfseries 05}
  (2020) 150} [\href{https://arxiv.org/abs/2002.11729}{{\ttfamily
  2002.11729}}].

\bibitem{Eberhardt:2019qcl}
L.~Eberhardt and M.R.~Gaberdiel, \emph{{String theory on AdS$_3$ and the
  symmetric orbifold of Liouville theory}},
  \href{https://doi.org/10.1016/j.nuclphysb.2019.114774}{\emph{Nucl. Phys. B}
  {\bfseries 948} (2019) 114774}
  [\href{https://arxiv.org/abs/1903.00421}{{\ttfamily 1903.00421}}].

\bibitem{Dei:2019osr}
A.~Dei, L.~Eberhardt and M.R.~Gaberdiel, \emph{{Three-point functions in
  AdS$_{3}$/CFT$_{2}$ holography}},
  \href{https://doi.org/10.1007/JHEP12(2019)012}{\emph{JHEP} {\bfseries 12}
  (2019) 012} [\href{https://arxiv.org/abs/1907.13144}{{\ttfamily
  1907.13144}}].

\bibitem{Schwimmer:1986mf}
A.~Schwimmer and N.~Seiberg, \emph{{Comments on the N=2, N=3, N=4
  Superconformal Algebras in Two-Dimensions}},
  \href{https://doi.org/10.1016/0370-2693(87)90566-1}{\emph{Phys. Lett. B}
  {\bfseries 184} (1987) 191}.

\bibitem{Sevrin:1988ew}
A.~Sevrin, W.~Troost and A.~Van~Proeyen, \emph{{Superconformal Algebras in
  Two-Dimensions with N=4}},
  \href{https://doi.org/10.1016/0370-2693(88)90645-4}{\emph{Phys. Lett. B}
  {\bfseries 208} (1988) 447}.

\bibitem{Gaberdiel:2015mra}
M.R.~Gaberdiel and R.~Gopakumar, \emph{{Stringy Symmetries and the Higher Spin
  Square}}, \href{https://doi.org/10.1088/1751-8113/48/18/185402}{\emph{J.
  Phys. A} {\bfseries 48} (2015) 185402}
  [\href{https://arxiv.org/abs/1501.07236}{{\ttfamily 1501.07236}}].

\bibitem{Guo:2021uiu}
B.~Guo and S.~Hampton, \emph{{Partial Spectral Flow in the D1D5 CFT}},
  \href{https://arxiv.org/abs/2112.10573}{{\ttfamily 2112.10573}}.

\bibitem{Lunin:2001fv}
O.~Lunin and S.D.~Mathur, \emph{{Metric of the multiply wound rotating
  string}}, \href{https://doi.org/10.1016/S0550-3213(01)00321-2}{\emph{Nucl.
  Phys. B} {\bfseries 610} (2001) 49}
  [\href{https://arxiv.org/abs/hep-th/0105136}{{\ttfamily hep-th/0105136}}].

\bibitem{Lunin:2000yv}
O.~Lunin and S.D.~Mathur, \emph{{Correlation functions for M**N / S(N)
  orbifolds}}, \href{https://doi.org/10.1007/s002200100431}{\emph{Commun. Math.
  Phys.} {\bfseries 219} (2001) 399}
  [\href{https://arxiv.org/abs/hep-th/0006196}{{\ttfamily hep-th/0006196}}].

\bibitem{Pakman:2009ab}
A.~Pakman, L.~Rastelli and S.S.~Razamat, \emph{{Extremal Correlators and
  Hurwitz Numbers in Symmetric Product Orbifolds}},
  \href{https://doi.org/10.1103/PhysRevD.80.086009}{\emph{Phys. Rev. D}
  {\bfseries 80} (2009) 086009}
  [\href{https://arxiv.org/abs/0905.3451}{{\ttfamily 0905.3451}}].

\bibitem{Pakman:2009mi}
A.~Pakman, L.~Rastelli and S.S.~Razamat, \emph{{A Spin Chain for the Symmetric
  Product CFT(2)}}, \href{https://doi.org/10.1007/JHEP05(2010)099}{\emph{JHEP}
  {\bfseries 05} (2010) 099} [\href{https://arxiv.org/abs/0912.0959}{{\ttfamily
  0912.0959}}].

\bibitem{Pakman:2009zz}
A.~Pakman, L.~Rastelli and S.S.~Razamat, \emph{{Diagrams for Symmetric Product
  Orbifolds}}, \href{https://doi.org/10.1088/1126-6708/2009/10/034}{\emph{JHEP}
  {\bfseries 10} (2009) 034} [\href{https://arxiv.org/abs/0905.3448}{{\ttfamily
  0905.3448}}].

\bibitem{Burrington:2012yn}
B.A.~Burrington, A.W.~Peet and I.G.~Zadeh, \emph{{Twist-nontwist correlators in
  $M^N/S_N$ orbifold CFTs}},
  \href{https://doi.org/10.1103/PhysRevD.87.106008}{\emph{Phys. Rev. D}
  {\bfseries 87} (2013) 106008}
  [\href{https://arxiv.org/abs/1211.6689}{{\ttfamily 1211.6689}}].

\bibitem{Burrington:2012yq}
B.A.~Burrington, A.W.~Peet and I.G.~Zadeh, \emph{{Operator mixing for string
  states in the D1-D5 CFT near the orbifold point}},
  \href{https://doi.org/10.1103/PhysRevD.87.106001}{\emph{Phys. Rev. D}
  {\bfseries 87} (2013) 106001}
  [\href{https://arxiv.org/abs/1211.6699}{{\ttfamily 1211.6699}}].

\bibitem{Roumpedakis:2018tdb}
K.~Roumpedakis, \emph{{Comments on the S$_{N}$ orbifold CFT in the large
  $N$-limit}}, \href{https://doi.org/10.1007/JHEP07(2018)038}{\emph{JHEP}
  {\bfseries 07} (2018) 038}
  [\href{https://arxiv.org/abs/1804.03207}{{\ttfamily 1804.03207}}].

\bibitem{GarciaiTormo:2018vqv}
J.~Garcia~i Tormo and M.~Taylor, \emph{{Correlation functions in the D1-D5
  orbifold CFT}}, \href{https://doi.org/10.1007/JHEP06(2018)012}{\emph{JHEP}
  {\bfseries 06} (2018) 012}
  [\href{https://arxiv.org/abs/1804.10205}{{\ttfamily 1804.10205}}].

\bibitem{Burrington:2018upk}
B.A.~Burrington, I.T.~Jardine and A.W.~Peet, \emph{{The OPE of bare twist
  operators in bosonic $S_N$ orbifold CFTs at large $N$}},
  \href{https://doi.org/10.1007/JHEP08(2018)202}{\emph{JHEP} {\bfseries 08}
  (2018) 202} [\href{https://arxiv.org/abs/1804.01562}{{\ttfamily
  1804.01562}}].

\bibitem{Dei:2019iym}
A.~Dei and L.~Eberhardt, \emph{{Correlators of the symmetric product
  orbifold}}, \href{https://doi.org/10.1007/JHEP01(2020)108}{\emph{JHEP}
  {\bfseries 01} (2020) 108}
  [\href{https://arxiv.org/abs/1911.08485}{{\ttfamily 1911.08485}}].

\bibitem{Guo:2019pzk}
B.~Guo and S.D.~Mathur, \emph{{Lifting of states in 2-dimensional $N = 4$
  supersymmetric CFTs}},
  \href{https://doi.org/10.1007/JHEP10(2019)155}{\emph{JHEP} {\bfseries 10}
  (2019) 155} [\href{https://arxiv.org/abs/1905.11923}{{\ttfamily
  1905.11923}}].

\end{thebibliography}\endgroup
\end{document}